\documentclass[11pt]{article}
\usepackage[T1]{fontenc}
\usepackage[utf8]{inputenc}
\usepackage{cite,float,url,amsmath,amssymb,nicefrac,epsfig}
 
\usepackage[english]{babel}
\usepackage[absolute]{textpos}
\usepackage{graphicx}
\usepackage{color}
\usepackage{mathrsfs}    
\usepackage{slashed}      
\usepackage{framed}
\usepackage{booktabs}

\usepackage{multirow}
\makeatletter

\@addtoreset{equation}{section}
\makeatother

\definecolor{shadecolor}{rgb}{1,0.8,0.3}

\begin{document}

\begin{titlepage}
\renewcommand{\thefootnote}{\fnsymbol{footnote}}
\setcounter{footnote}{0}

\vspace*{-2cm}
\begin{flushright}
CFTP/12-004 \\
 PCCF RI 12-03\\

\par\end{flushright}

\begin{flushright}
\vspace*{2mm}
\par\end{flushright}

\begin{center}
\vspace*{15mm}
 \textbf{\Large Revisiting the $\Gamma\left(K\rightarrow e\nu\right)/
\Gamma\left(K\rightarrow\mu\nu\right)$
ratio in }

\vspace*{2mm}
 \textbf{\Large supersymmetric unified models}\\

\vspace*{1cm}
\textbf{ Renato M. Fonseca$^{a}$, J. C. Rom\~ao$^{a}$ and A. M. Teixeira$^{b}$}
\par
\end{center}

\begin{center}
\vspace*{0.5cm}

\par\end{center}

\begin{center}
$^{a}$ Centro de F\'{\i}sica Te\'orica de Part\'{\i}culas, CFTP,
Instituto Superior 
T\'ecnico, \\
Universidade T\'ecnica de Lisboa, Av. Rovisco Pais 1, 1049-001 Lisboa,
Portugal
\par\end{center}

\begin{center}
\vspace*{0.2cm}
 $^{b}$ Laboratoire de Physique Corpusculaire, CNRS/IN2P3 -- UMR
6533,\\
 Campus des C\'ezeaux, 24 Av. des Landais, F-63171 Aubi\`ere Cedex, France
\par\end{center}

\vspace*{10mm}
 
\begin{abstract}
It has been pointed out that supersymmetric extensions of the Standard
Model can induce significant changes to the theoretical prediction of
the ratio $\Gamma\left(K\rightarrow e\nu\right)/\Gamma
\left(K\rightarrow\mu\nu\right)\equiv R_{K}$, through lepton flavour
violating couplings. In this work we carry out a full computation of
all one-loop corrections to the relevant $\nu\ell H^{+}$ vertex, and
discuss the new contributions to $R_{K}$ arising in the context of
different constrained (minimal supergravity inspired) models which
succeed in accounting for neutrino data, further
considering the possibility of accommodating a near future observation
of a $\mu\to e\gamma$ transition. We also re-evaluate the prospects
for $R_{K}$ in the framework of unconstrained supersymmetric
models. In all cases, we address the question of whether it is
possible to saturate the current experimental sensitivity on $R_{K}$
while in agreement with the recent limits on $B$-meson
decay observables (in particular BR($B_{s}\to\mu^+\mu^-$) and
BR($B_{u}\to\tau\nu$)), as well as BR($\tau\to e
\gamma$) and available collider constraints.  Our findings reveal
that in view of the recent bounds, and even when enhanced by effective
sources of flavour violation in the right-handed
$\tilde{e}-\tilde{\tau}$ sector, constrained supersymmetric (seesaw)
models typically provide excessively small contributions to
$R_{K}$. Larger contributions can be found in more general settings, 
where the charged Higgs mass can be effectively lowered, and even
further enhanced  in the unconstrained MSSM.
However, our analysis clearly shows that even in this last case
SUSY contributions to $R_{K}$ are still unable to
saturate the current experimental bounds on this observable, especially 
due to a strong tension with the $B_{u}\to\tau\nu$ bound.

\end{abstract}
\vspace*{3mm}
 {\footnotesize KEYWORDS: Supersymmetry, neutrinos, meson decays,
flavour violation}{\footnotesize \par}

\end{titlepage}

\section{Introduction}

Neutrino oscillations have provided the first experimental manifestation
of flavour violation in the lepton sector, fuelling the need to consider
extensions of the Standard Model (SM) that succeed in explaining the
smallness of neutrino masses and the observed pattern of their 
mixings~\cite{Fogli:2011qn,Schwetz:2011zk,theta13}.
In addition to the many facilities dedicated to study neutral leptons,
there is currently a great experimental effort to search for signals
of flavour violation in the charged lepton sector (cLFV), since such
an observation would provide clear evidence for the existence of new
physics beyond the SM (trivially extended to accommodate massive neutrinos).
The quest for the origin of the underlying mechanism of flavour violation
in the lepton sector has been actively pursued in recent years, becoming
even more challenging as the MEG experiment is continually improving
the sensitivity to $\mu\to e\gamma$ decays~\cite{arXiv:1107.5547},
thus opening the door for a possible measurement (observation) in
the very near future. The current bounds on other radiative decays
(i.e. $\ell_{i}\to\ell_{j}\gamma$), or three-body decays ($\ell_{i}\to3\ell_{j}$)
are already impressive~\cite{PDG}, and are expected to be further
improved in the future.

Supersymmetric (SUSY) extensions of the SM offer new sources of CP
and flavour violation, in both quark and lepton sectors. Given the
strong experimental constraints, especially on CP and flavour violating
observables involving the strongly interacting sector, phenomenological
analyses in general favour the so-called ``flavour-blind'' mechanisms
of SUSY breaking, where universality of the soft breaking terms is
assumed at some high energy scale: in these constrained scenarios,
the only sources of flavour violation (FV) are the quark and charged
lepton Yukawa couplings. In order to accommodate current neutrino
data, mechanisms of neutrino mass generation, such as the seesaw (in
its different realisations - for a review of the latter, 
see for instance~\cite{Abada:2007ux,Abada:2011rg}),
are often implemented in the framework of (constrained) SUSY models:
in the case of the so-called ``SUSY-seesaw'', radiatively induced
flavour violation in the slepton sector~\cite{Borzumati:1986qx}
can provide sizable contributions to cLFV observables. The latter
have been extensively studied, both at high- and low-energies, over
the past years (see e.g.~\cite{Raidal:2008jk}).
Flavour violation can be also incorporated in a more
phenomenological approach, where at low-energies new sources of FV
are present in the soft SUSY breaking terms. However, these are
severely constrained by a large number of observables (see, 
e.g.~\cite{Antonelli:2009ws} and references therein).

In addition to the above mentioned rare lepton decays, leptonic and
semi-leptonic meson decays also offer a rich testing ground for cLFV.
Here we will be particularly interested in leptonic $K$ decays, which
(as is also the case of leptonic $\pi$ decays) constitute very good
probes of violation of lepton universality. The
potential of these observables, especially regarding SUSY extensions
of the SM, was firstly noticed in~\cite{Masiero:2005wr}, and later
investigated in greater detail 
in~\cite{Masiero:2008cb,Ellis:2008st,Girrbach:2012km}.

By themselves, these decays are heavily hampered by hadronic uncertainties
and, in order to reduce the latter (and render these decays an efficient
probe of new physics), one usually considers the ratio 
\begin{equation}
R_{K}\,\equiv\,\frac{\Gamma\left(K^{+}\rightarrow e^{+}\nu
\left[\gamma\right]\right)}{\Gamma\left(K^{+}\rightarrow\mu^{+}\nu
\left[\gamma\right]\right)}\,,\label{eq:rk:def}
\end{equation}
since in this case the hadronic uncertainties cancel to a very good
approximation. As a consequence, the SM prediction can be computed
with high precision~\cite{Marciano:1993sh,Finkemeier:1995gi,Cirigliano:2007xi}.
The most recent analysis has provided the following 
value~\cite{Cirigliano:2007xi}:
\begin{equation}
R_{K}^{\textrm{SM}}=(2.477\pm0.001)\times10^{-5}\,.\label{eq:Cirigliano:2007xi}
\end{equation}
On the experimental side, the NA62 collaboration has recently obtained
very stringent bounds~\cite{Goudzowski:2011tc}: 
\begin{equation}
R_{K}^{\textrm{exp}}\,=\,(2.488\pm0.010)\,\times10^{-5}\,,\label{eq:rk:NA62}
\end{equation}
 which should be compared with the SM prediction 
(Eq.~(\ref{eq:Cirigliano:2007xi})).
In order to do so, it is often useful to introduce the following
parametrisation, 
\begin{equation}
R_{K}^{\textrm{exp}}\,=\, R_{K}^{\textrm{SM}}\left(1+\Delta
  r\right)\,,
\quad\quad\Delta r\equiv\nicefrac{R_{K}}{R_{K}^{\textrm{SM}}}-1\,,
\label{eq:deltark}
\end{equation}
 where $\Delta r$ is a quantity denoting potential contributions
arising from scenarios of new physics (NP). Comparing the theoretical
SM prediction to the current bounds (i.e., Eqs.~(\ref{eq:Cirigliano:2007xi},
\ref{eq:rk:NA62})), one verifies that observation is compatible with
the SM (at 1$\sigma$) for 
\begin{equation}
\Delta r\,=\,\left(4\pm4\right)\times10^{-3}\,.\label{eq:deltarexp}
\end{equation}

Previous analyses have investigated supersymmetric contributions to
$R_{K}$ in different frameworks, as for instance low-energy SUSY
extensions of the SM (i.e. the unconstrained Minimal Supersymmetric
Standard Model (MSSM))~\cite{Masiero:2005wr,Masiero:2008cb,Girrbach:2012km},
or non-minimal grand unified models (where higher dimensional terms
contribute to fermion masses)~\cite{Ellis:2008st}. These studies
have also considered the interplay of $R_{K}$ with other important
low-energy flavour observables, magnetic and electric lepton moments
and potential implications for leptonic CP violation. Distinct computations,
based on an approximate parametrisation of flavour violating effects
- the Mass Insertion Approximation (MIA)~\cite{Hall:1985dx} - allowed
to establish that SUSY LFV contributions can induce large contributions
to the breaking of lepton universality, as parametrised by $\Delta r$.
The dominant FV contributions are in general associated to charged-Higgs
mediated processes, being enhanced due to non-holomorphic effects
- the so-called {}``HRS'' mechanism~\cite{Hall:1993gn} -, and
require flavour violation in the $RR$ block of the charged slepton
mass matrix. It is important to notice that these Higgs contributions
have been known to have an impact on numerous observables, and can
become especially relevant for the large $\tan\beta$ regime~\cite{Hou:1992sy,Hall:1993gn,Chankowski:1994ds,Babu:1999hn,Carena:1999py,Babu:2002et,Brignole:2003iv,Brignole:2004ah,Arganda:2004bz,Paradisi:2005tk,Paradisi:2006jp,RamseyMusolf:2007yb}.

In the present work, we re-evaluate the potential of a broad class
of supersymmetric extensions of the SM to saturate the
current measurement of $R_{K}$.
Contrary to previous studies, we conduct a full computation of the
one-loop corrections to the $\nu\ell H^{+}$ vertex,
taking into account the important
contributions from non-holormophic effective Higgs-mediated interactions.
When possible we establish a bridge between our results and approximate
analytical expressions in the literature, and we stress the potential
enhancements to the total SUSY contributions.
In our numerical analysis we re-investigate the
prospects regarding $R_{K}$
of a constrained MSSM onto which several seesaw realisations
are embedded (type
I~\cite{seesaw:I} and II~\cite{seesaw:II}, as well as the inverse
seesaw~\cite{Mohapatra:1986bd}), also briefly addressing
$L$--$R$ symmetric models~\cite{LRmodels1,LRmodels2}.
We then consider
more relaxed scenarios, such as non-universal Higgs mass (NUHM) models
at high-scale (which are known to enhance this 
class of observables~\cite{Ellis:2008st} due to 
potentially lighter charged Higgs boson masses),
and discuss the general prospects of unconstrained low-energy SUSY models.
In all cases, we revisit the $R_{K}$ observable in the light
of new experimental data: in addition to LHC bounds%
\footnote{
In our numerical analysis we do not require the
lightest Higgs to be in strict agreement with recent LHC 
search results~\cite{LHC:Higgs:2012}:
while in the general the case (especially for constrained (seesaw)
models), we only favour regimes where its mass is larger than 118
GeV, when considering semi-constrained and unconstrained models, a
significant part of the studied region does indeed comply with $m_{h}\sim125$
GeV.} on the sparticle spectrum~\cite{LHC:2011} and a number of low-energy
flavour-related bounds~\cite{PDG,arXiv:1107.5547}, we implement
the very recent LHCb results concerning the 
BR($B_{s}\to\mu^{+}\mu^{-}$)~\cite{Aaij:2012ac}.
As we discuss here, the increasing tension with low-energy
observables, in particular with $B_{u}\to\tau\nu$,
precludes sizable SUSY contributions to $R_{K}$ even
in the context of otherwise favoured candidate models as is the case
of semi-constrained and unconstrained SUSY models.

This document is organised as follows. Section~\ref{sec:RK:formulae}
is devoted to the computation of the 1-loop MSSM prediction for $R_{K}$.
We compare our (full) result to the approximations in the literature
by means of the mass insertion approximation (among other simplifications),
and discuss the dominant sources of flavour violation, and the implications
to other observables. Our results for a number of models are collected
in Section~\ref{sec:res}. Further discussion and concluding remarks
are given in Section~\ref{sec:concs}. In the Appendices, we detail
the computation of the renormalised charged lepton - neutrino - charged
Higgs vertex, and summarise the key features of two supersymmetric
seesaw realisations (types I and II) used in the numerical analysis.

\section{Supersymmetric contributions to $R_{K}$}

\label{sec:RK:formulae}

In the SM, the decay widths of pseudoscalar mesons into light leptons
are given by 
\begin{equation}
\Gamma^{\text{SM}}(P^{\pm}\!\!\to\ell^{\pm}\nu)=
\frac{G_{F}^{2}m_{P}m_{\ell}^{2}}{8\pi}
\left(\!1-\!\frac{m_{\ell}^{2}}{m_{P}^{2}}\right)^{2}\!\! f_{P}^{2}
|V_{qq^{\prime}}|^{2} ,
\label{eq:SM:Pdecays}
\end{equation}
 where $P$ denotes $\pi,K,D$ or $B$ mesons, with mass $m_{P}$
and decay constant $f_{P}$, and where $G_{F}$ is the Fermi constant,
$m_{\ell}$ the lepton mass and $V_{qq^{\prime}}$ the corresponding
Cabibbo-Kobayashi-Maskawa (CKM) matrix element. These decays are helicity
suppressed (as can be seen from the factor $m_{\ell}^{2}$ in 
Eq.~(\ref{eq:SM:Pdecays})),
and the prediction for their amplitude is thus hampered by the hadronic
uncertainties in the meson decay constants. As mentioned in the Introduction,
ratios of these amplitudes are independent of $f_{P}$ to a very good
approximation, and the SM prediction can then be computed very precisely.
Concerning the kaon decay ratio $R_{K}$, the SM prediction (inclusive
of internal bremsstrahlung radiation) is~\cite{Cirigliano:2007xi}
\begin{equation}
R_{K}^{\text{SM}}\,=\,\left(\frac{m_{e}}{m_{\mu}}\right)^{2}\,
\left(\frac{m_{K}^{2}-m_{e}^{2}}{m_{K}^{2}-m_{\mu}^{2}}\right)^{2}\,
\left(1+\delta R_{\text{QED}}\right)\,,
\end{equation}
 where $\delta R_{\text{QED}}$ is a small electromagnetic correction
accounting for internal bremsstrahlung and structure-depen\-dent effects
($\delta R_{\text{QED}}=(-3.60\pm0.04)\%$~\cite{Cirigliano:2007xi}).

\begin{figure}[!htb]
\begin{centering}
\begin{tabular}{c}
\includegraphics[clip,width=50mm]{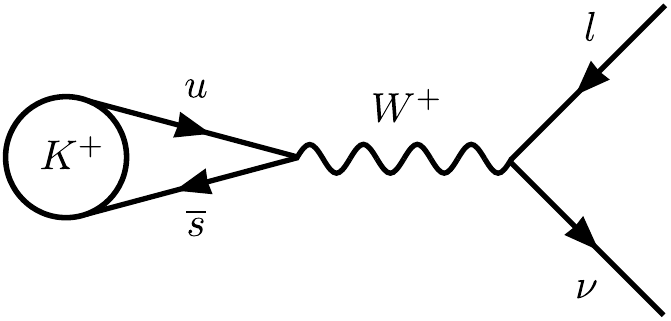}\\[+2mm]  
\includegraphics[clip,width=50mm]{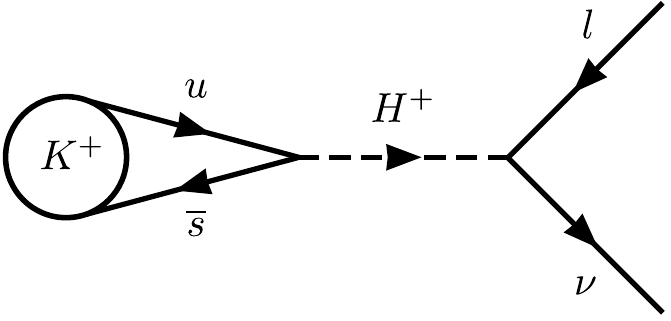}
\\
\end{tabular}
\par\end{centering}
\caption{Tree level contributions to $R_{K}$ - SM and charged Higgs.}
\label{fig:RK:SM:Higgs} 
\end{figure}

In supersymmetric models, the extended Higgs sector can play an important
r\^ole in lepton flavour violating transitions and decays (see~\cite{Hou:1992sy,Hall:1993gn,Chankowski:1994ds,Babu:1999hn,Carena:1999py,Babu:2002et,Brignole:2003iv,Brignole:2004ah,Arganda:2004bz,Paradisi:2005tk,Paradisi:2006jp,RamseyMusolf:2007yb}).
The effects of the additional Higgs are also sizable in meson decays
through a charged Higgs boson, as schematically depicted in 
Fig.~\ref{fig:RK:SM:Higgs}.
In particular, for kaons, one finds~\cite{Hou:1992sy} 
\begin{align}
\Gamma(K^{\pm}\to\ell^{\pm}\nu)\,=&\,\Gamma^{\text{SM}}(K^{\pm}
\to\ell^{\pm}\nu) \nonumber\\
&\times 
\left(1-\tan^{2}\beta\,\frac{m_{K}^{2}}{m_{H^+}^{2}}\,
\frac{m_{s}}{m_{s}+m_{u}}\right)^{2};\label{eq:kaon:gamma:smsusy}
\end{align}
however, despite this new tree-level contribution, $R_{K}$ is unaffected,
as the extra factor does not depend on the (flavoured) leptonic part
of the process.

New contributions to $R_{K}$ only emerge at higher order: at one-loop
level, there are box and vertex contributions, wave function renormalisation,
which can be both lepton flavour conserving (LFC) and lepton flavour
violating. Flavour conserving contributions arise from loop corrections
to the $W^{\pm}$ propagator, through heavy Higgs exchange (neutral
or charged) as well as from chargino/neu\-tralino-sleptons (in the latter
case stemming from non-universal slepton masses, i.e., a selectron-smuon
mass splitting). As concluded in~\cite{Masiero:2005wr}, in the framework
of SUSY models where lepton flavour is conserved, the new contributions
to $\Delta r^{\text{SUSY}}$ are too small to be within experimental
reach.

On the other hand, Higgs mediated LFV processes are capable of providing
an important contribution when the kaon decays into a electron plus
a tau-neutrino. For such LFV Higgs couplings to arise, the leptonic
doublet ($L$) must couple to more than one Higgs doublet. However,
at tree level in the MSSM, $L$ can only couple to $H_{1}$, and therefore
such LFV Higgs couplings arise only at loop level, due to the generation
of an effective non-holomorphic coupling between $L$ and $H_{2}^{*}$
- the HRS mechanism~\cite{Hall:1993gn} - which is a crucial ingredient
in enhancing the Higgs contributions to LFV observables. In what follows,
we address the impact of these non-holomorphic terms for $R_{K}$.

\subsection{LFV Higgs mediated contributions to $R_{K}$}

We consider as starting point the MSSM, defined by its superpotential
and soft-SUSY breaking Lagrangian. We detail below the relevant terms
for our discussion:

\begin{equation}
\mathcal{W}=\hat{U}^{c} Y^{u} \hat{Q} \hat{H}_{2}
-\hat{D}^{c}  Y^{d} \hat{Q} \hat{H}_{1} - \hat{E}^{c}  
Y^{l} \hat{L} \hat{H}_{1} -\mu \hat{H}_{1} \hat{H}_{2} ,
\label{eq:Wlepton0:def}
\end{equation}
\begin{align}
\mathcal{V}_{\text{soft}}=&-\mathcal{L}_{\text{soft}}\, 
 =(M_{\alpha}\,\psi_{\alpha}\,\psi_{\alpha}+\text{h.c.})+
m_{H_{i}}^{2}\, H_{i}^{*}\, H_{i}
\nonumber \\
&+(B\, H_{1}\, H_{2}+\text{h.c.})
+\tilde{\ell}_{L}^{*}\, m_{\tilde{L}}^{2}\,
\tilde{\ell}_{L}\,+\tilde{\ell}_{R}^{*}\, 
m_{\tilde{R}}^{2}\tilde{\ell}_{R}\,
\nonumber \\
 & 
+\,(H_{1}\,\tilde{\ell}_{R}^{*}\, 
A^{l}\,\tilde{\ell}_{L}\,+\text{h.c.})+...\,,
\end{align}
 where $M_{\alpha}$ denotes the soft-gaugino mass terms, {}``...''
stand for the squark terms, and we have omitted flavour indices. For
the SU(2) superfield products, we adopt the convention 
$\hat{H}_{1}\,\hat{H}_{2}\equiv\hat{H}_{1}^{1}\,
\hat{H}_{2}^{2}-\hat{H}_{1}^{2}\,\hat{H}_{2}^{1}$
(and likewise for similar cases).

\begin{figure}[!htb]
\begin{centering}
\includegraphics[clip, width=85mm]{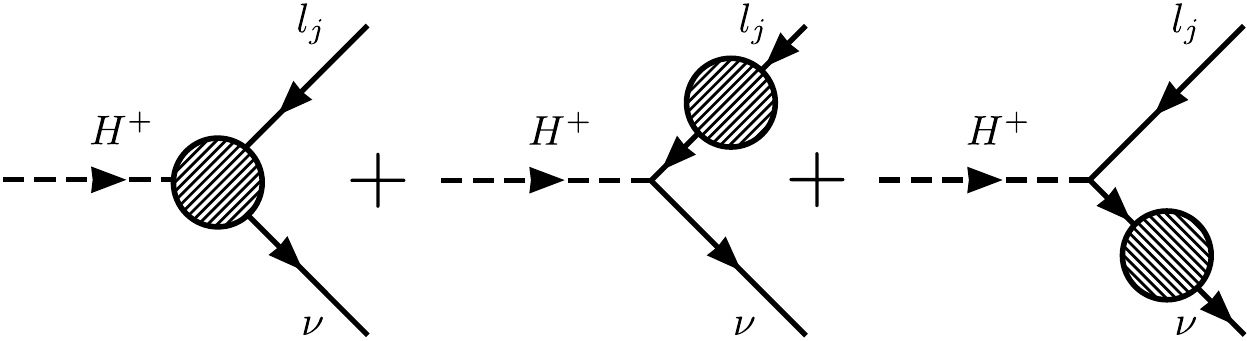}
\par\end{centering}
\caption{Corrections to the $\nu\ell H^{+}$ vertex, as discussed in the text.}
\label{fig:RK:Hvertex} 
\end{figure}

\bigskip{}
From an effective theory approach, the HRS mechanism can be accounted
for by additional terms, corresponding to the higher-order corrections
to the Higgs-neutrino-charged lepton interaction (schematically depicted
in Fig.~\ref{fig:RK:Hvertex}). 

At tree-level, the Lagrangian describing the $\nu\ell H^{\pm}$ interaction
is given by 
\begin{align}
\mathcal{L}_{0}^{H^{\pm}}\,=&\,\overline{\nu}_{L}\,
Y^{l\dagger}\,\ell_{R}\, H_{1}^{-*}+\textrm{h.c.}\,
\nonumber\\
=&\,\left(2^{3/4}\, G_{F}^{1/2}\right)\,\tan\beta\,
\overline{\nu}_{L}\, M^{l}\,\ell_{R}\, H^{+}+\textrm{h.c.}\,,
\label{eq:L:nulH0}
\end{align}
with $M^{l}=\textrm{diag}\left(m_{e},m_{\mu},m_{\tau}\right)$. At
loop level, two new terms are generated: 
$\overline{\nu}_{L}\Delta^{+}\ell_{R}H_{2}^{+}-
\overline{\ell}_{L}\Delta^{0}\ell_{R}H_{2}^{0}+\textrm{h.c.}$.
The second one, with $\Delta^{0}$, forces a redefinition of the charged
lepton Yukawa couplings, $Y^{l\dagger}=\frac{M^{l}}{v_{1}}$ $\to$
$Y^{l\dagger}\approx\frac{M^{l}}{v_{1}}-\Delta^{0}\tan\beta$, which
in turn implies a redefinition of the charged lepton propagator; the
term with $\Delta^{+}$ corrects the Higgs-neutrino-charged lepton
vertex%
\footnote{An extensive discussion on the radiatively induced couplings which
are at the origin of the HRS effect can be found in~\cite{Borzumati:1999sp}.%
}. Once these terms are taken into account, the interaction Lagrangian,
Eq.~(\ref{eq:L:nulH0}), becomes 
\begin{align}
\mathcal{L}^{H^{\pm}} & \,=\,\left(2^{3/4}\, G_{F}^{1/2}\right)\,
\tan\beta\,\,\overline{\nu}_{L}\, M^{l}\,\ell_{R}\, H^{+}\,
\nonumber\\
&+\,\cos\beta\,\,\overline{\nu}_{L}\,\left(\Delta^{+}
-\Delta^{0}\,\tan^{2}\beta\right)\,\ell_{R}\, H^{+}+\textrm{h.c.}\,.
\label{eq:L:nulH}
\end{align}
Since in the $\textrm{SU(2)}_{L}$-preserving limit $\Delta^{+}=\Delta^{0}$,
it is reasonable to assume that, after electroweak (EW) symmetry breaking,
both terms remain approximately of the same order of magnitude. Hence,
it is clear that the contribution associated with $\Delta^{0}$ (the
loop contribution to the charged lepton mass term) will be enhanced
by a a factor of $\tan^{2}\beta$ when compared to the one associated
with $\Delta^{+}$. This simple discussion allows to understand the
origin of the dominant SUSY contribution%
\footnote{\label{footpag5}There are additional corrections to the 
$\overline{q}q^{\prime}H^{\pm}$
vertex, which are mainly due to a similar modification of the the
quark Yukawa couplings - especially that of the strange quarks. This
amounts to a small multiplicative effect on $\Delta r$ which we will
not discuss here (see~\cite{Girrbach:2012km} for details).%
} to $R_{K}$.

As we proceed to discuss, a quantitative assessment of the corrections
to $\Delta^{+}$ and $\Delta^{0}$ requires considering the higher-order
effects on the vertex $\overline{\nu}_{L}\, Z^{H}\,\ell_{R}\, H^{+}$
(see also~\cite{Bellazzini:2010gn}). The $Z^{H}$ matrix depends
on the following (loop-induced) quantities: 
\begin{itemize}
\item $\eta_{L}^{\ell}$ and $\eta_{L}^{\nu}$ (corrections to the kinetic
terms of $\ell_{L}$ and $\nu_{L}$); 
\item $\eta_{m}^{\ell}$ (correction to the charged lepton mass term); 
\item $\eta^{H}$ (correction to the $\nu\ell H$ vertex). 
\end{itemize}
The expressions for the distinct $\eta$-parameters can be found in
Appendix~\ref{sec:app1}. Instead of $Z^{H}$, which includes both
tree and loop level effects, it proves to be more convenient to use
the following combination, 
\begin{equation}
-\frac{\tan\beta}{2^{3/4}G_{F}^{1/2}}\,
\left(\frac{m_{K}}{m_{H^+}}\right)^{2}\,\frac{m_{s}}{m_{s}+m_{u}}\, 
Z^{H}\,\left(M^{l}\right)^{-1}\,\equiv\,\epsilon\,\mathbf{1}+
\Delta\,,\label{eq:epsilondelta}
\end{equation}
where 
\begin{align}
\epsilon & =-\tan^{2}\beta\,
\left(\frac{m_{K}}{m_{H^+}}\right)^{2}\frac{m_{s}}{m_{s}+m_{u}}\,,
\label{eq:epsilon:def}\\
\Delta & =\epsilon\,\left[\frac{\eta_{L}^{\ell}}{2}-
\frac{\eta_{L}^{\nu}}{2}+\left(\frac{\eta^{H}}{2^{3/4}\, 
G_{F}^{1/2} \tan\beta}-\eta_{m}^{\ell}\right)
\left(M^{l}\right)^{-1}\right] .\label{eq:delta:def}
\end{align}
In the above, $\epsilon$ encodes the tree level Higgs mediated amplitude
(which does not change the SM prediction for $R_{K}$), while $\Delta$,
a matrix in lepton flavour space, encodes the 1-loop effects. The
main contribution is expected to arise from $\eta_{m}^{\ell}$.

The $\Delta r$ observable is then related to $\epsilon$ and $\Delta$
as follows: 
\begin{equation}
\Delta r\equiv\frac{R_{K}}{R_{K}^{\text{SM}}}-1\,
=\,\frac{\left[\left(\mathbf{1}+
\frac{\Delta^{\dagger}}{1+\epsilon}\right)
\left(\mathbf{1}+\frac{\Delta}{1+\epsilon}\right)\right]_{ee}}
{\left[\left(\mathbf{1}+\frac{\Delta^{\dagger}}{1+\epsilon}\right)
\left(\mathbf{1}+\frac{\Delta}{1+\epsilon}\right)\right]_{\mu\mu}}-1\,.
\label{eq:deltar:epsilondelta}
\end{equation}
If the slepton mixing is sufficiently large, this expression can be
approximated as 
\begin{equation}
\Delta r\,\approx\,2\,\operatorname{Re}(\Delta_{ee})\,
+\,(\Delta^{\dagger}\Delta)_{ee}\,.
\end{equation}
In the above, the first (linear) term on the right hand-side is due
to an interference with the SM process, and is thus lepton flavour
conserving. As shown in~\cite{Masiero:2005wr}, this contribution
can be enhanced through both large $RR$ and $LL$ slepton mixing.
On the other hand, the quadratic term $(\Delta^{\dagger}\Delta)_{ee}$
can be augmented mainly through a large LFV contribution from $\Delta_{\tau e}$,
which can only be obtained in the presence of significant $RR$ slepton
mixing.

\subsection{Generating $\Delta r$: sources of flavour violation and experimental
constraints}

In order to understand the dependence of $\Delta r$ on the SUSY parameters,
and the origin of the dominant contributions to this observable, an
approximate expression for $\Delta$ is required. Firstly, we notice
that the previous discussion, leading to Eq.~(\ref{eq:L:nulH}),
suggests that the $\eta_{m}^{\ell}$ term is responsible for the dominant
contributions to $\Delta r$. Thus, in what follows, and for the purpose
of obtaining simple analytical expressions, we shall neglect the contributions
of the other terms (although these are included in the numerical analysis
of Section~\ref{sec:res}). A fairly simple analytical insight can
be obtained when working in the limit in which the virtual particles
in the loops (sleptons and gauginos) are assumed to have similar masses,
so that their relative mass splittings are indeed small. In this limit,
one can Taylor-expand the loop functions entering $\eta_{m}^{\ell}$
(see Appendix~\ref{sec:app1}); working to third order in this expansion,
and keeping only the terms enhanced by a factor of 
$m_{\tau}\,\tan\beta\,\frac{m_{\text{SUSY}}}{m_{\text{EW}}}$
(where $m_{\text{SUSY}},\, m_{\text{EW}}$ denote the SUSY breaking
scale and EW scale, respectively), we obtain 
\begin{align}
\Delta r\,\sim\, & \left[1+X\left(1-
\frac{9}{10}\frac{\delta}{\overline{m}_{\widetilde{\ell},\chi^{0}}^{2}}\right)
\left(m_{\tilde{L}}^{2}\right)_{e\tau}\right]^{2}-1
\nonumber\\
&+X^{2}\left[-\mu^{2}
+\delta\left(3-\frac{3}{10}\frac{\mu^{2}+2M_{1}^{2}}
{\overline{m}_{\widetilde{\ell},\chi^{0}}^{2}}\right)\right]^{2},
\label{eq:deltar:approx:long}
\end{align}
where $\mu$, $M_{1}$ and $(m_{\tilde{L}}^{2})_{e\tau}$ denote the
low-energy values of the Higgs bilinear term, bino soft-breaking mass,
and off-diagonal entry of the soft-breaking left-handed slepton mass
matrix, respectively. We have also introduced
$\overline{m}_{\widetilde{\ell},\chi^{0}}^{2}=
\frac{1}{2}(\langle{m}_{\widetilde{\ell}}^{2}\rangle+
\langle{m}_{\chi^{0}}^{2}\rangle)$, the average mass squared of
sleptons and neutralinos ($\approx m^2_{\text{SUSY}}$), and
$\delta=\frac{1}{2}(\langle{m}_{\widetilde{\ell}}^{2}\rangle-
\langle{m}_{\chi^{0}}^{2}\rangle)$, the corresponding splitting. The
quantity $X$ is given by
\begin{equation}
X\,\equiv\,\frac{1}{192\pi^{2}}\, m_{K}^{2}\, 
g'^{2}\,\mu\, M_{1}\,\frac{\tan^{3}\beta}{m_{H^+}^{2}}\,
\frac{m_{\tau}}{m_{e}}\,
\frac{\left(m_{\tilde{R}}^{2}\right)_{\tau e}}
{(\overline{m}_{\widetilde{\ell},\chi^{0}}^{2})^{3}}\,,
\label{eq:deltar:approx:X}
\end{equation}
and it illustrates in a transparent (albeit approximate) way the origin
of the terms contributing to the enhancement of $R_{K}$: in addition
to the factor ${\tan^{3}\beta}/{m_{H^+}^{2}}$, usually associated
with Higgs exchanges, the crucial flavour violating source emerges
from the off-diagonal $(\tau e)$ entry of the right-handed slepton
soft-breaking mass matrix.

Using the above analytical approximation, one easily recovers the
results in the literature, usually obtained using the MIA. For instance,
Eq.~(11) of Ref.~\cite{Masiero:2005wr} amounts to 
\begin{equation}
\Delta r\,\sim\,2X\,\left(m_{\tilde{L}}^{2}\right)_{e\tau}\,
+\, X^{2}\,\left(m_{\tilde{L}}^{2}\right)_{e\tau}^{2}\,
+\, X^{2}\,\delta^{2}\,,
\label{eq:deltar:approx}
\end{equation}
which stems from having kept the dominant (crucial) second and third
order contributions in the expansion: 
$X^{2}\delta^{2}$ and $2X\left(m_{\tilde{L}}^{2}\right)_{e\tau}+
X^{2}\left(m_{\tilde{L}}^{2}\right)_{e\tau}^{2}$,
respectively.

\bigskip{}
Regardless of the approximation considered, it is thus clear that
the LFV effects on kaon decays into a $e\nu$ or $\mu\nu$ pair can
be enhanced in the large $\tan\beta$ regime (especially in the presence
of low values of $m_{H^+}$), and via a large $RR$ slepton mixing
$\left(m_{\tilde{R}}^{2}\right)_{\tau e}$. Although the latter is
indeed the privileged source, notice that, as can be seen from
 Eq.~(\ref{eq:deltar:approx}),
a strong enhancement can be obtained from sizable flavour violating
entries of the left-handed slepton soft-breaking mass, 
$\left(m_{\tilde{L}}^{2}\right)_{e\tau}$.
This is in fact a globally flavour conserving effect (which can also
account for negative contributions to $R_{K}$). Previous experimental
measurements of $R_{K}$ appeared to favour values smaller than the
SM theoretical estimation, thus motivating the study of regimes leading
to negative values of $\Delta r$~\cite{Masiero:2005wr}, 
but these regimes have now become disfavoured in view of the present
bounds, Eq.~(\ref{eq:deltarexp}).

Clearly, these Higgs mediated exchanges, as well as the FV terms at
the origin of the strong enhancement to $R_{K}$, will have an impact
on a number of other low-energy observables, as can be easily inferred
from the structure of Eqs.~(\ref{eq:deltar:approx:long}-\ref{eq:deltar:approx}).
This has been extensively addressed in the 
literature~\cite{Masiero:2005wr,Masiero:2008cb,Ellis:2008st,Girrbach:2012km},
and here we will only briefly discuss the most relevant observables:
electroweak precision data on the anomalous electric and magnetic
moments of the electron, as well as the naturalness of the electron
mass, directly constrain the $\eta_{m}^{\ell}$ corrections (and $\eta_{L}^{\ell}$,
$\eta^{H}$); low-energy cLFV observables, such as $\tau\to\ell\gamma$
and $\tau\to3\ell$ decays are also extremely sensitive probes of
Higgs mediated exchanges, and in the case of $\tau-e$ transitions,
depend on the same flavour violating entries. It has been suggested
that positive and negative values of $\Delta r$ can be of the order
of 1\%, still in agreement with data on the electron's electric dipole
moment and on 
$\tau\rightarrow\ell\gamma$~\cite{Masiero:2005wr,Masiero:2008cb,Ellis:2008st}.
Finally, other meson decays, such as $B\to\ell\ell$ 
(and $B\to\ell\nu$), exhibit a similar dependence on $\tan\beta$,
$\tan^{n}\beta/{m_{H^+}}^{4}$~\cite{BToLL} ($n$ ranging from 2 to
6, depending on the other SUSY parameters), 
and may also lead to indirect bounds on $\Delta r$. In particular,
the strict bounds on BR($B_{u}\to\tau\nu$)~\cite{PDG}
and the very recent limits on BR($B_{s}\to\mu^{+}\mu^{-}$)~\cite{Aaij:2012ac}
might severely constrain the allowed regions in SUSY parameter space
for large $\tan\beta$. Although we will come to this issue in greater
detail when discussing the numerical results, it is clear that the
similar nature of the $K^{+}\to\ell\nu$ and $B_{u}\to\tau\nu$
processes (easily inferred
from a generalization of Eq.~(\ref{eq:kaon:gamma:smsusy}), see
e.g.~\cite{Hou:1992sy,Isidori:2006pk}) 
will lead to a tension when light charged Higgs masses are considered
to saturate the bounds on $R_{K}$.

Supersymmetric models of neutrino mass generation (such as the SUSY
seesaw) naturally induce sizable cLFV contributions, via radiatively
generated off-diagonal terms in the $LL$ (and to a lesser extent
$LR$) slepton soft-breaking mass matrices~\cite{Borzumati:1986qx}.
In addition to explaining neutrino masses and mixings, such models
can also easily account for values of BR($\mu\to e\gamma$), within
the reach of the MEG experiment. In view of the recent confirmation
of a large value for the Chooz angle 
($\theta_{13}\sim8.8^{\circ}$)~\cite{theta13}
and on the impact it might have on $(m_{\tilde{L}}^{2})_{e\tau}$,
in the numerical analysis of the following section we will also consider
different realisations of the SUSY seesaw (type I~\cite{seesaw:I},
II~\cite{seesaw:II} and inverse~\cite{Mohapatra:1986bd}), embedded
in the framework of constrained SUSY models. We
will also revisit semi-constrained scenarios allowing for light values
of $m_{H^+}$, re-evaluating the predictions for $R_{K}$ under
a full, one loop-computation, and in view of recent experimental data. Finally,
we confront these (semi-)constrained scenarios with general, low-energy
realisations, of the MSSM.

\section{Prospects for $R_{K}$: unified vs unconstrained SUSY models}
\label{sec:res}

In this section we evaluate the SUSY contributions to $R_{K}$, with
the results obtained via the full expressions for $\Delta r$, as
described in Section~\ref{sec:RK:formulae}. These were implemented
into the SPheno public code~\cite{Porod:2003um}, which was accordingly
modified to allow the different studies. It is important to stress
that although some approximations have still been done (as previously
discussed), the results based on the present computation strongly
improve upon those so far reported in the literature (mostly obtained
using the MIA). Although the different contributions cannot be easily
disentangled due to having carried a full computation, our results
automatically include \textit{all} one-loop lepton flavour violating
and lepton flavour conserving contributions (in association with charged
Higgs mediation, see footnote~\ref{footpag5}). As mentioned before,
we evaluate $R_{K}$ in the framework of constrained,
semi-constrained (NUHM) and unconstrained SUSY models. Concerning
the first two,  we assume some flavour blind mechanism of SUSY
breaking (for instance minimal supergravity (mSUGRA) inspired), so
that the soft breaking parameters obey universality conditions at
some high-energy scale, which we choose to be the gauge coupling unification
scale $M_{\text{GUT}}\sim10^{16}$ GeV, 
\begin{align}
 \left(m_{\tilde{Q}}\right)_{ij}^{2}\,
=&\,\left(m_{\tilde{U}}\right)_{ij}^{2}\,
=\,\left(m_{\tilde{D}}\right)_{ij}^{2}\,
=\,\left(m_{\tilde{L}}\right)_{ij}^{2}\,
=\,
\left(m_{\tilde{R}}\right)_{ij}^{2}
\nonumber \\
=&\, m_{0}^{2}\,\delta_{ij}\,,\,\,\nonumber \\[+2mm]
 \left(A^{l}\right)_{ij}\,=&\, A_{0}\,(Y^{l})_{ij}\,.
\label{eq:msugra.univ}
\end{align}
In the above, $m_{0}$ and $A_{0}$ are the universal scalar soft-breaking
mass and trilinear couplings of the cMSSM, and $i,j$ denote lepton
flavour indices ($i,j=1,2,3$). In the latter case, the gaugino masses
are also assumed to be universal, their common value being denoted
by $M_{1/2}$. We will also consider the supersymmetrisation of several
mechanisms for neutrino mass generation. More specifically, we have
considered the type I and type II SUSY seesaw 
(as detailed in Appendix~\ref{app:seesaw}).
We briefly comment on the inverse SUSY seesaw, and discuss a $L-R$
model.

The strict universality boundary conditions of
Eqs.~(\ref{eq:msugra.univ}) will be relaxed for the Higgs sector
when we address NUHM scenarios, so that in the latter case we will have
\begin{align}
 & m_{H_{1}}^{2}\,\neq m_{H_{2}}^{2}\,\neq\, m_{0}^{2}\,.\label{eq:msugra.nuhm}
\end{align}
All the above universality hypothesis will be further relaxed when, for completeness,
and to allow a final comparison with previous analyses, we address
the low-energy unconstrained MSSM.

\bigskip{}
In our numerical analysis, we took into account LHC bounds on the
SUSY spectrum~\cite{LHC:2011}, as well as the constraints from low-energy
flavour dedicated experiments~\cite{PDG}, and 
neutrino data~\cite{Fogli:2011qn,Schwetz:2011zk}.
In particular, concerning lepton flavour violation, 
we have considered~\cite{PDG,arXiv:1107.5547}:
\begin{align}
 & \text{BR}(\tau\to e\gamma)\,<3.3\times10^{-8}\,
\quad(90\%\text{C.L.})\,,\label{eq:cLFVbounds1}\\
 & \text{BR}(\tau\to3\, e)\,<2.7\times10^{-8}\,
\quad(90\%\text{C.L.})\,,\label{eq:cLFVbounds2}\\
 & \text{BR}(\mu\to e\gamma)\,<2.4\times10^{-12}\,
\quad(90\%\text{C.L.})\,,\\
 & \text{BR}(B_{u}\to\tau\nu)\,>9.7\times10^{-5}\,
\quad(2\,\sigma)\,.
\label{eq:cLFVbounds3}
\end{align}
Also relevant are the recent LHCb bounds~\cite{Aaij:2012ac} 
\begin{align}
 & \text{BR}(B_{s}\to\mu^{+}\mu^{-})\,<4.5\times10^{-9}\,
\quad(95\%\text{C.L.})\,,\label{eq:Bbounds1}\\
 & \text{BR}(B\to\mu^{+}\mu^{-})\,<1.03\times10^{-9}\,
\quad(95\%\text{C.L.})\,.\label{eq:Bbounds2}
\end{align}

When addressing models for neutrino mass generation, we take the following
(best-fit) values for the neutrino mixing angles~\cite{Schwetz:2011zk}
(where $\theta_{13}$ is already in good agreement with the most recent
results from~\cite{theta13}), 
\begin{align}
 & \sin^{2}\theta_{12}\,=\
 0.312_{-0.015}^{+0.017},\quad\sin^{2}\theta_{23}\,=\
 0.52_{-0.07}^{+0.06},
\nonumber \\[+2mm]
&\sin^{2}\theta_{13}\approx0.013_{-0.005}^{+0.007}\,, 
\label{eq:mixingangles:data}\\[+2mm]    
 & \Delta\,
 m_{\text{12}}^{2}\,=\,(7.59_{-0.18}^{+0.20})\,\times10^{-5}\,\,\text{eV}^{2}\,,
\\[+2mm]    
 &\Delta\,
 m_{\text{13}}^{2}\,=(2.50_{-0.16}^{+0.09})\,\times10^{-3}\,\,\text{eV}^{2}\,. 
\end{align}
Regarding the leptonic mixing matrix ($U_{\text{MNS}}$) we adopt
the standard parametrisation. In the present analysis, all CP violating
phases are set to zero%
\footnote{We will assume that we are in a strictly CP conserving framework,
where all terms are taken to be real. This implies that there will
be no contributions to observables such as electric dipole moments,
or CP asymmetries.%
}.

\subsection{mSUGRA inspired scenarios: cMSSM and the 
SUSY seesaw}

We begin by re-evaluating, through a full computation of the one-loop
corrections, the maximal amount of supersymmetric contributions to
$R_{K}$ in constrained SUSY scenarios. For a first evaluation of
$R_{K}$, we consider different cMSSM (mSUGRA-like) points, defined in
Table~\ref{table:mSUGRA:points}.  Among them are several cMSSM
benchmark points from \cite{AbdusSalam:2011fc}, representative of low
and large $\tan\beta$ regimes, as well as some variations.  Notice
that, as mentioned before, these choices are compatible with having a
Higgs boson mass above 118 GeV but will be excluded once we require
$m_{h}$ to lie close to 125 GeV as suggested by LHC
results~\cite{LHC:Higgs:2012}.

\begin{table*}
\begin{centering}
\begin{tabular}{cccccc}\hline
 & $m_{0}$ (GeV)  & $M_{1/2}$ (GeV)  & $\tan\beta$  & 
$A_{0}$ (GeV)  & sign($\mu$) \\
\hline
10.3.1  & 300  & 450  & 10  & 0  & 1\\
P20  & 330  & 500  & 20  & -500 & 1\\
P30  & 330  & 500  & 30  & -500  & 1\\
40.1.1  & 330  & 500  & 40  & -500  & 1\\
40.3.1  & 1000  & 350  & 40  & -500  & 1\\
\hline
\end{tabular}
\par\end{centering}
\caption{cMSSM (benchmark) points used in the numerical analysis.}
\label{table:mSUGRA:points} 
\end{table*}

As could be expected from
Eqs.~(\ref{eq:deltar:approx:long}-\ref{eq:deltar:approx}), in a strict
cMSSM scenario (in agreement with the experimental bounds above
referred to) the SUSY contributions to $R_{K}$ are extremely small;
motivated by the need to accommodate neutrino data, and at the same
time accounting for values of BR($\mu\to e\gamma$) within MEG reach,
we implement type I and type II seesaws in mSUGRA-inspired models (see
Appendices~\ref{app:seesawI} and~\ref{app:seesawII}).  Regarding the
heavy-scale mediators, we considered degenerate right-handed
neutrinos, as well as degenerate scalar triplets. We set the seesaw
scale aiming at maximising the (low-energy) non-diagonal entries of
the soft-breaking slepton mass matrices, while still in agreement with
the current low-energy bounds (see
Eqs.~(\ref{eq:cLFVbounds1}-\ref{eq:Bbounds2})).  In particular, we
have tried to maximise the $LL$ contributions to $\Delta r$, i.e.,
$(m_{\tilde{L}}^{2})_{e\tau}$, and to obtain BR($\mu\to e\gamma$)
within MEG reach (i.e.  $10^{-13}\lesssim$ BR($\mu\to
e\gamma$)$\lesssim2.4\times10^{-12}$).  However, and due to the fact
that both seesaw realisations fail to account for radiatively induced
LFV in the right-handed slepton sector, one finds values $|\Delta
r|\lesssim2\times10^{-8}$.  It is worth emphasising that if one
further requires $m_{h}$ to lie close to 125 GeV (as suggested by
recent findings~\cite{LHC:Higgs:2012}), then one is led to regions in
mSUGRA parameter space where, due to the much heavier sparticle masses
and typically lower values of $\tan\beta$, the SUSY contributions to
$R_{K}$ become even further suppressed.

Thus, and even under a full computation of the corrections to the
$\nu\ell H^{+}$ vertex, we nevertheless confirm
that, as firstly put forward in the 
analyses of~\cite{Masiero:2005wr,Masiero:2008cb}
strictly constrained SUSY and SUSY seesaw models indeed fail to account
for values of $R_{K}$ close to the present limits.

\bigskip{}
Clearly, new sources of flavour violation, associated to the right-handed
sector are required: in what follows, we maintain universality of
soft-breaking terms allowing, at the grand unified (GUT) scale, for
a single $\tau-e$ flavour violating entry in $m_{\tilde{R}}^{2}$.
This approach is somewhat closer to the lines 
of~\cite{Masiero:2005wr,Masiero:2008cb,Ellis:2008st,Girrbach:2012km},
although in our computation we will still conduct a full evaluation
of the distinct contributions to $\Delta r$, and we consider otherwise
universal soft-breaking terms. Without invoking a specific framework/scenario
of SUSY breaking that would account for such a pattern, we thus set
\begin{equation}
\delta_{31}^{RR}\,=\,\frac{(m_{\tilde{R}}^{2})_{\tau e}}{m_{0}^{2}}\,
\neq0\,.
\label{eq:mia:sleptons}
\end{equation}
As discussed above, low-energy constraints on LFV observables (especially
$\tau\rightarrow e\gamma$), severely constrain this entry.

\begin{figure*}[htb]
\centering
\begin{tabular}{cc}
\includegraphics[width=0.42\linewidth]{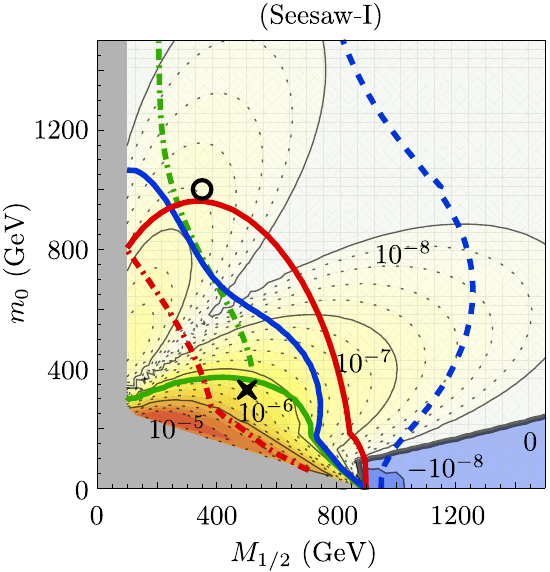}\hspace*{5mm}  &
\hspace*{5mm}  
\includegraphics[width=0.42\linewidth]{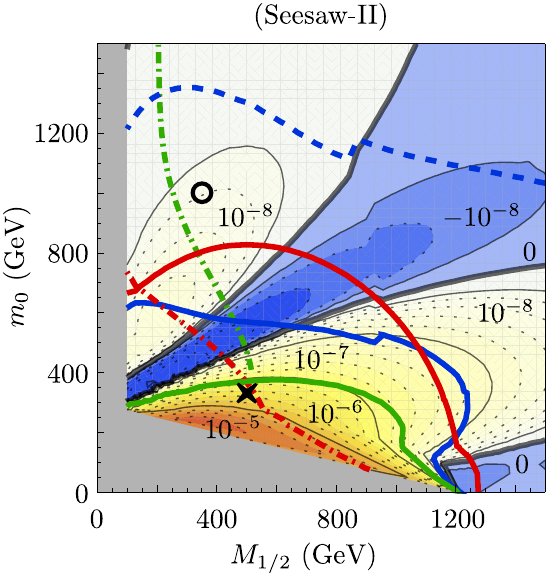} 
\\ 
\end{tabular}
\caption{$m_{0}-M_{1/2}$ plane for $\tan\beta=40$ and $A_{0}=-500$
GeV, with $\delta_{31}^{RR}=0.1$. On the left (right) panel, a type I
(II) SUSY seesaw, considering degenerate heavy mediators. Contour
lines denote values of $\Delta r$ (decreasing values: positive - in
association with an orange-yellow-white colour gradient; negative -
blue gradients); solid (grey) regions are excluded due to the
- requirement of having the correct EWSB. A green dot-dashed line
corresponds to the present LHC bounds on the
cMSSM~\cite{LHC-Limits}. A full green line delimits the BR($\tau\to
e\gamma$) exclusion region, while full (dot-dashed) red lines
correspond to the bounds on BR($B_{s}\to\mu^{+}\mu^{-}$)
(BR($B_{u}\to\tau\nu$)). Finally, the region delimited by blue lines
corresponds to having BR($\mu\to e\gamma$) within MEG reach (current
bound - solid line, future sensitivity - dashed line). Crosses
(circles) correspond to the benchmark point 40.1.1
(40.3.1).} 
\label{fig:cMSSM:seesaw} 
\end{figure*}

In Fig.~\ref{fig:cMSSM:seesaw},
we present our results for $\Delta r$ scanning the $m_{0}-M_{1/2}$
plane for a regime of large $\tan\beta$. We have set $\delta_{31}^{RR}=0.1$,
$\tan\beta=40$, and taken $A_{0}=-500$ GeV. The surveys displayed
in the panels correspond to having embedded a type I (left) or type
II (right) seesaw onto this near-mSUGRA framework. 

As can be readily seen from Fig.~\ref{fig:cMSSM:seesaw}, once the
constraints from low-energy observables have been applied, in the
type I SUSY seesaw, the maximum values for 
$\Delta r$ are $\mathcal{O}(10^{-7})$,
associated to the region with a lighter SUSY spectra (which
is in turn disfavoured by a ``heavy'' light Higgs). Even for
the comparatively small non-universality, $\delta_{31}^{RR}=0.1$,
a considerable region of the parameter space is excluded due to excessive
contributions to BR($B_{u}\to\tau\nu$) and BR($\tau\to e\gamma$), thus
precluding the possibility of large values of $\Delta r$. 
In a regime of large $\tan\beta$, the contributions to
BR($B_{s}\to\mu^{+}\mu^{-}$) are also sizable, and the recent LHCb
results seem to exclude the regions of the parameter space where
one could still have $\Delta r\sim\mathcal{O}(10^{-6,-7})$. 
The excessive SUSY
contributions to BR($B_{s}\to\mu^{+}\mu^{-}$) can be somewhat 
reduced by
adjusting $A_0$ (in Fig.~\ref{fig:cMSSM:seesaw} we fixed $A_0=-500$
GeV) and the values of  $\Delta r$
can be slightly augmented by increasing $\delta_{31}^{RR}$; in the
latter case, the $\tau\to e\gamma$ bound proves to be the most
constraining, and  values of $\Delta
r$ larger than $\mathcal{O}(10^{-6,-7})$ cannot be obtained in these
constrained SUSY seesaw models.

The situation is somewhat different for the type II case: firstly
notice that a sizable region in the $m_{0}-M_{1/2}$ plane is associated
to negative contributions to $R_{K}$, which are currently disfavoured.
In the remaining (allowed) parameter space, the values of $\Delta r$
are slightly smaller than for the type I case: this is a consequence
of a non trivial interplay between a smaller value for the splitting
$\delta=\frac{1}{2}(\langle{m}_{\widetilde{\ell}}^{2}\rangle-
\langle{m}_{\chi^{0}}^{2}\rangle)$
(induced by a lighter spectra), and a lighter charged Higgs boson.
(We notice that accommodating light neutral Higgs with $m_{h}>118$
GeV is also comparatively more difficult in the type II SUSY seesaw.)

Notice that in both SUSY seesaws it is fairly easy to accommodate
a potential observation of BR($\mu\to e\gamma$) $\sim10^{-13}$ by
MEG, taking for instance $M_{{\rm Seesaw}}\sim10^{12}$ GeV for the
type I and II seesaw mechanisms.

For both cases, larger values of $\delta_{31}^{RR}=0.5$ can be taken,
but these typically lead to conflicting situations with low-energy
observables; lowering $\tan\beta$ can ease the existing tension,
at the expense of also reducing $\Delta r$. We summarise this on
Table~\ref{table:mSUGRA:results}, for simplicity in association
with a type I SUSY seesaw.

\begin{table*}[htb]
\begin{center}
\begin{tabular}{lccccccc}
\hline
 & {\small $\delta_{31}^{RR}$ } & {\small $\Delta r$ } & {\small }%
\begin{tabular}{{@{}c@{}}}
{\small $m_{H^+}$}\\
{\small (GeV)}\\
\end{tabular} & {\small BR($\tau\to e\gamma$) } & {\small }%
\begin{tabular}{{@{}c@{}}}
{\small BR($B_{u}\to\tau\nu$)}\\
{\small $(\times10^{-4})$}\\
\end{tabular} & {\small }%
\begin{tabular}{{@{}c@{}}}
{\small BR($B_{s}\to\mu^{+}\mu^{-}$) }\\
{\small $(\times10^{-9})$}\\
\end{tabular} & {\small BR($\mu\to e\gamma$)}\\[+1mm]
\hline\\[-3mm]
{\small 10.3.1 - I } & {\small 0 } & {\small 7.2$\times10^{-11}$} &
{\small 715 } & {\small 2.5$\times10^{-16}$} & {\small 1.17} & {\small
  4.0} & {\small 7.2$\times10^{-14}$}
\\
{\small 10.3.1 - I } & {\small 0.1 } & {\small 8.5$\times10^{-11}$} &
{\small 715} & {\small 2.9$\times10^{-10}$} & {\small 1.17} & {\small
  4.0} & {\small 1.8$\times10^{-13}$ }
\\
{\small 10.3.1 - I } & {\small 0.5 } & {\small 5.1$\times10^{-9}$} &
{\small 715 } & {\small 8.5$\times10^{-9}$ } & {\small 1.12} & {\small
  4.0} & {\small 9.7$\times10^{-15}$}
\\
\hline\\[-3mm]
{\small \hskip 0.1mm P20 - I } & {\small 0.1 } & {\small 4.3$\times10^{-9}$} &
{\small 800} & {\small 3.5$\times10^{-9}$ } & {\small 1.15} & {\small
  4.0} & {\small 2.0$\times10^{-12}$}
\\[+1mm]
\hline\\[-3mm]
{\small \hskip 0.1mm P30 - I } & {\small 0.1 } & {\small 1.2$\times10^{-7}$} &
{\small 725 } & {\small 1.4$\times10^{-8}$} & {\small 1.11} & {\small
  4.3} & {\small 1.7$\times10^{-14}$ }
\\[+1mm]
\hline\\[-3mm]
{\small 40.3.1 - I } & {\small 0 } & {\small 1.6$\times10^{-8}$} &
{\small 818} & {\small 3.1$\times10^{-15}$ } & 1.09 & {\small 4.4} &
{\small 1.2$\times10^{-12}$}
\\
{\small 40.3.1 - I } & {\small 0.1 } & {\small 6.0$\times10^{-8}$} &
{\small 818} & {\small 2.9$\times10^{-10}$ } & 1.09 & {\small 4.4} &
{\small 1.2$\times10^{-12}$}
\\
{\small 40.3.1 - I } & {\small 0.5 } & {\small 2.0$\times10^{-6}$} &
{\small 818} & {\small 2.0$\times10^{-8}$ } & 1.09 & {\small 4.4} &
{\small 3.3$\times10^{-12}$}
\\[+1mm]
\hline
\end{tabular}
\end{center}
\caption{$\Delta r$ and other low-energy observables 
for different mSUGRA
points, considering a type I seesaw, and distinct 
values of $\delta_{31}^{RR}$.
The values of the seesaw scale were varied from $1.3\times10^{12}$
GeV to $5\times10^{10}$ GeV, in order to comply with the limits/future
sensitivity on BR($\mu\to e\gamma$).}
\label{table:mSUGRA:results}
\end{table*}
A few comments are in order regarding the summary of
Table~\ref{table:mSUGRA:results}: even with a large value for
$\delta_{31}^{RR}$, and in the large $\tan\beta$ regime, the maximum
attainable values for $\Delta r$ are much below the current
experimental sensitivity, at most $2 \times 10^{-6}$. 
As mentioned before, if we further take 
into account the recent
discovery of a new boson at LHC~\cite{LHC:Higgs:2012}
with a mass around
125 GeV, and interpret it as the lightest neutral CP-even Higgs boson
of the MSSM, the attainable values for $\Delta r$ will be extremely
small.

\begin{figure*}[htb]
\centering
\begin{tabular}{cc}
\includegraphics[width=0.45\linewidth]{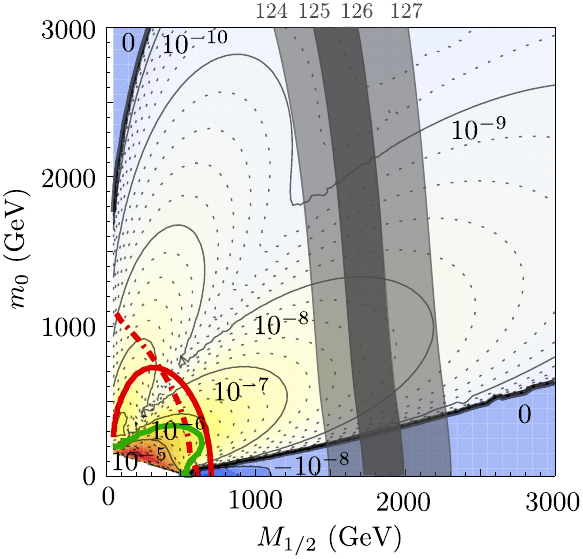} 
&\includegraphics[width=0.45\linewidth]{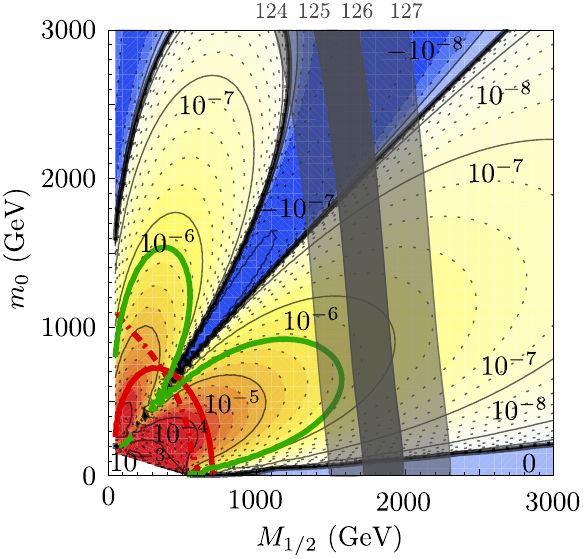} 
\end{tabular}
\caption{mSUGRA (type I seesaw) $m_{0}-M_{1/2}$ plane for
  $\tan\beta=40$ and $A_{0}=0$ 
GeV, with $\delta_{31}^{RR}=0.1$ (left panel) and $\delta_{31}^{RR}=0.7$
(right panel). Contour lines denote values of $\Delta r$ (decreasing
values: positive - in association with an orange-yellow-white colour
gradient; negative - blue gradients). A full green line delimits the
BR($\tau\to e\gamma$) exclusion region, while full (dot-dashed) red
lines correspond to the bounds on BR($B_{s}\to\mu^{+}\mu^{-}$)
(BR($B_u\to\tau\nu$)). 
Superimposed are the regions for the Higgs boson mass: the dark band
is for $125\leq m_{h^{0}}\leq126$ (GeV) and the lighter one marks
the region where $124\leq m_{h^{0}}\leq127$ (GeV).}
\label{fig:cMSSM:survey} 
\end{figure*}

In order to conclude this part of the analysis
we provide a comprehensive overview of
the constrained MSSM prospects regarding $R_{K}$, presenting in
Fig.~\ref{fig:cMSSM:survey} a survey of the (type I seesaw) mSUGRA
parameter space, for two different regimes of $\delta_{31}^{RR}$, 
taking \textit{all} present bounds (including the recent ones on 
$m_{h}$) into account. The panels of Fig.~\ref{fig:cMSSM:survey} allow to recover the
information that could be expected from the discussion following 
Fig.~\ref{fig:cMSSM:seesaw}: for fixed values of $A_0$ and $\tan
\beta$, increasing $\delta_{31}^{RR}$ indeed allows to augment the
SUSY contributions to $\Delta r$ although, as can be seen from the
right-panel, the constraints from BR($\tau\to e\gamma$) become
increasingly harder to accommodate.
(Notice that the latter could be avoided
by increasing the SUSY scale (i.e. on regions of the parameter space
with large $m_0$ and/or $M_{1/2}$) - however, and as visible from 
Fig.~\ref{fig:cMSSM:survey}, in a constrained SUSY framework this
would lead to heavier charged Higgs masses, and in turn to suppressed 
contributions to $\Delta r$.)

Although we do not display an analogous plot here, the situation is
very similar for the type II SUSY seesaw (slightly even more
constrained due to the fact that accommodating $m_{h}\sim125$ GeV is
more difficult in these models~\cite{Hirsch:2012ti}).

In view of the above discussion it is clear that even taking into
account all 1-loop corrections to the $\nu\ell H^{+}$ vertex, values
of $\Delta r$, large enough to saturate current observation, cannot be
reached in the framework of constrained SUSY models (and its seesaw
extensions accommodating neutrino data). In this sense, and even
though we have followed a different approach, our results follow the
conclusions of~\cite{Ellis:2008st}. We also stress that recent
experimental bounds (both from flavour observables and collider
searches) add even more severe constraints to the maximal possible
values of $\Delta r$.

\subsection{mSUGRA inspired scenarios: inverse seesaw 
and $L-R$ models }

We briefly comment here on the prospects of the inverse SUSY seesaw
concerning $R_{K}$: recently, it was pointed out that some flavour
violating observables can be enhanced by as much as two orders of
magnitude in a model with the inverse seesaw mechanism~\cite{Abada:2011hm}.
Within such a framework, right-handed (s)neutrino masses can be relatively
light, and as a consequence these ${\nu_{R}}$, ${\widetilde{\nu}_{R}}$
states do not decouple from the theory until the TeV scale, hence
potentially providing important contributions to different low-energy
processes. Nevertheless, the specific contributions to $\Delta r$
are suppressed by a factor $\frac{m_{e}^{2}}{m_{\tau}^{2}}$, with
respect to those discussed above (see Eq.~(\ref{eq:deltar:approx:X})),
so that we do not expect a significant enhancement of SUSY 1-loop effects
to $R_{K}$ due to the inverse seesaw mechanism.

For completeness (and although we do not provide specific details
here), we have considered a specific $L-R$ seesaw model~\cite{LRmodels2}.
In this framework, non-vanishing values of $\delta_{31}^{RR}$ can
be dynamically generated. We have numerically verified that typically
one finds $\delta_{31}^{RR}\lesssim0.01$ (we do not dismiss that
larger values might be found, although certainly requiring a considerable
amount of fine-tuning in the input parameters). We have not done a
dedicated $\Delta r$ calculation for this case, but taking into account
that the effect scales with $(\delta_{31}^{RR})^{2}$, we also expect
the typical range for $\Delta r$ to be far below the current experimental
sensitivity.

\subsection{mSUGRA inspired scenarios: NUHM}

As can be seen from the approximate expression
for $\Delta r$ in Eqs.~(\ref{eq:deltar:approx:X}, \ref{eq:deltar:approx}),
regimes associated with both large $\tan\beta$ and a light charged
Higgs can greatly enhance this observable~\cite{Ellis:2008st} 
($\Delta r\propto\nicefrac{\tan^{6}\beta}{m_{H^+}^{4}}$).
By relaxing the mSUGRA-inspired universality conditions for the Higgs
sector, as occurs in NUHM scenarios, one can indeed have very low
masses for the $H^{+}$ boson at low energies. This regime corresponds
to a narrow strip in parameter space where $m_{H_{1}}^{2}\!\approx\!
m_{H_{2}}^{2}$, in particular when both are close to 
$-(2.2\:\textrm{TeV})^{2}$.
In addition to favouring electroweak symmetry breaking, since
$m^2_{H^{+}}\sim\left|m_{H_{1}}^{2}-m_{H_{2}}^{2}\right|$
(even accounting for RG evolution of the parameters down to the weak
scale), it is expected that the charged Higgs can be made
very light with some fine tuning~\cite{Ellis:2008st}. 
In order to explore the maximal possible values
of $\Delta r$, a small scan was conducted around this region, 
where $m_{H^+}$ changes very rapidly (see
Table~\ref{tab:NUHM}).
\begin{table}
\centering
\begin{tabular}{cccccc}
\hline 
 & $m_{0}$  & $M_{1/2}$  & $m_{H_{1}}^{2}$, $m_{H_{2}}^{2}$  &
 $\tan\beta$  & $\delta_{31}^{RR}$ \\
& $\text{\footnotesize (GeV)}$  & $\text{\footnotesize (GeV)}$  & 
$\text{\footnotesize (GeV}^{2}\text{\footnotesize )}$
& &  
\\ 
\hline\\[-3mm]
Min  & 0 & 100  & $-5.2\times10^{6}$  & 40  & 0.1\\
Max  & 1500  & 1500  & $-4.6\times10^{6}$  & 40  & 0.7\\
\hline
\end{tabular}
\caption{Range of NUHM parameters leading to the scan
of Fig.~\ref{fig:NUHM:deltar}. }
\label{tab:NUHM} 
\end{table}
\begin{figure*}[!htb]
\centering 
\begin{tabular}{cc}
\includegraphics[width=0.49\linewidth]{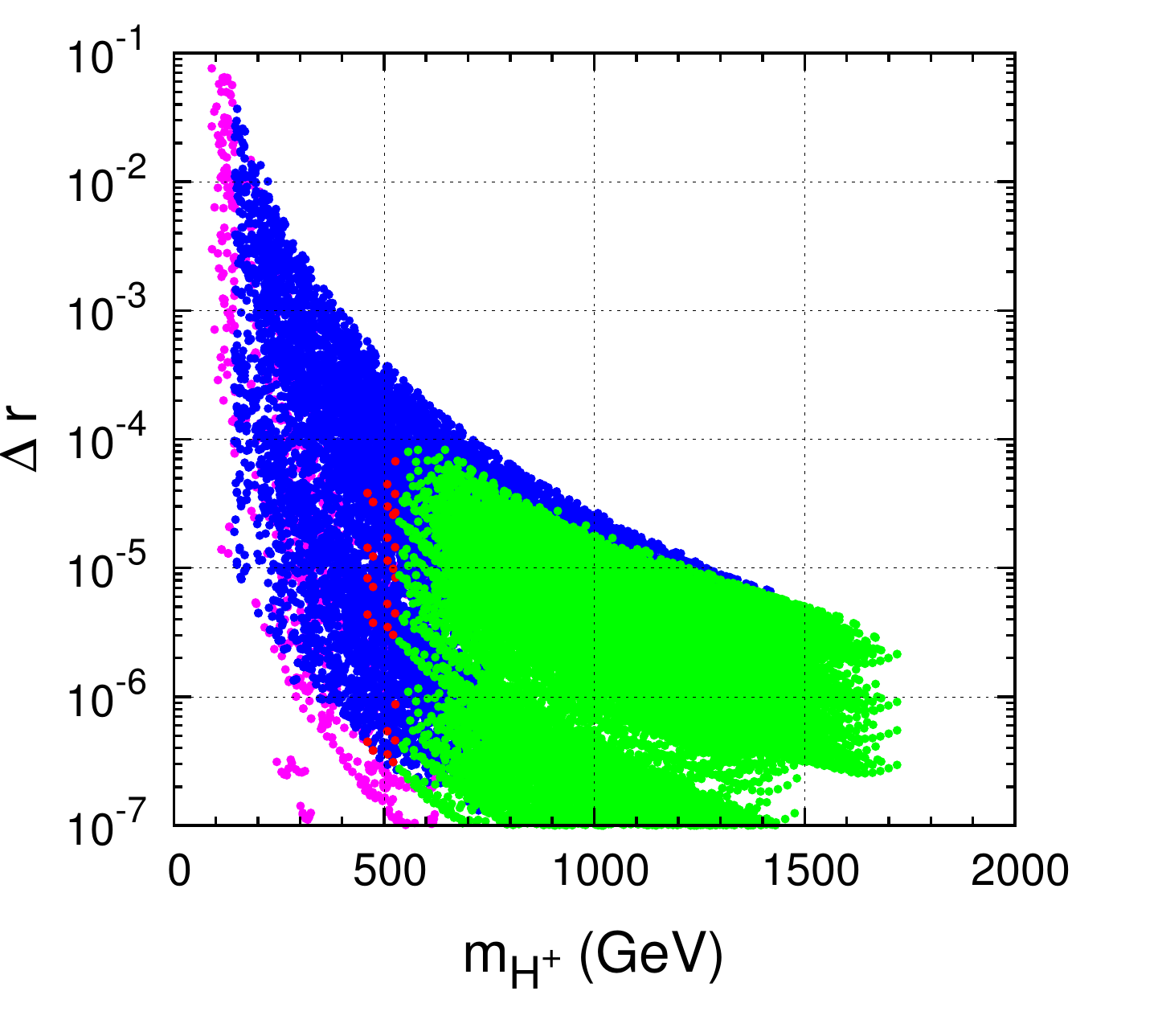}
&\includegraphics[width=0.49\linewidth]{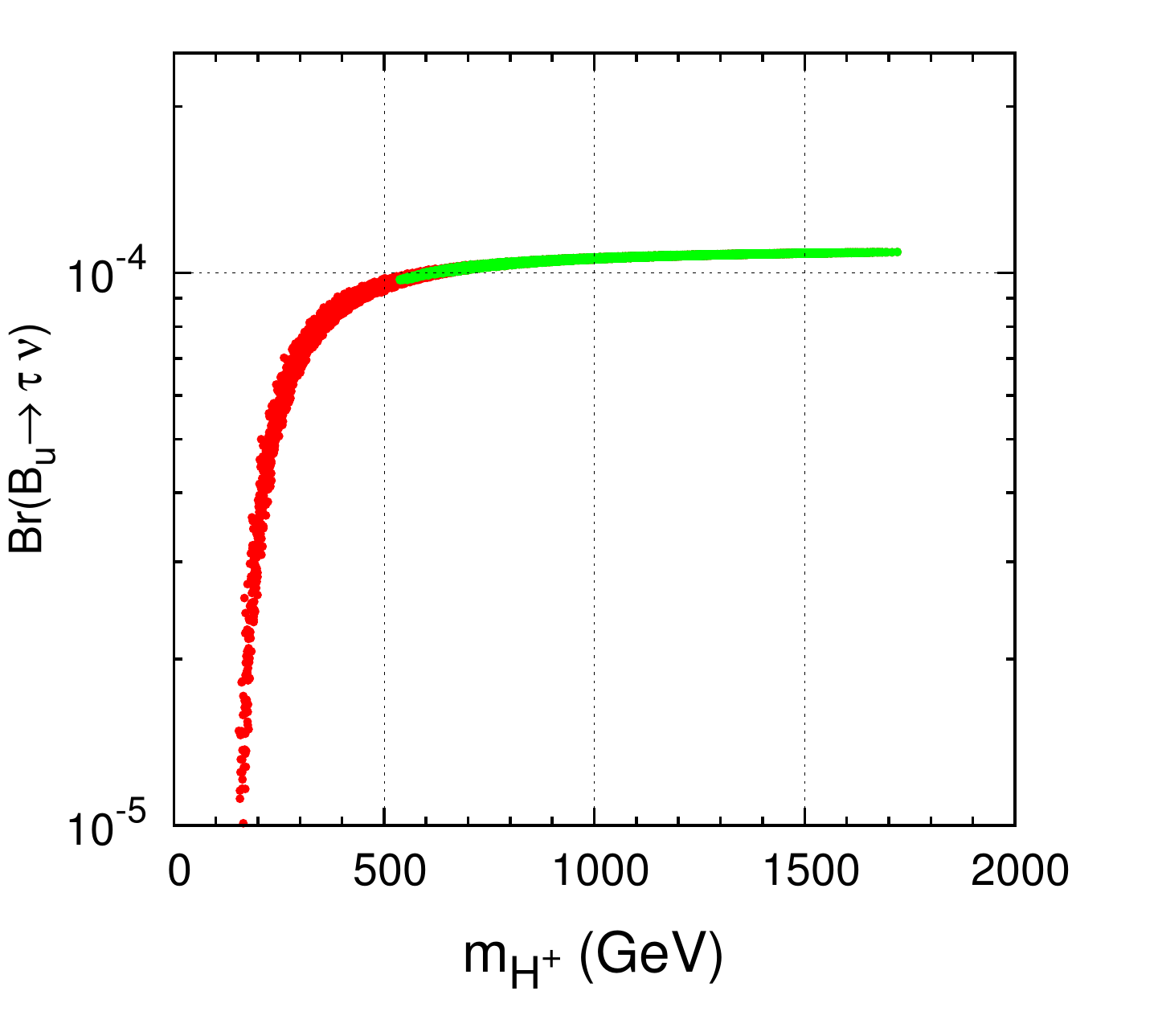}
\end{tabular}
\caption{Left panel:
$\Delta r$ as a function of the charged Higgs mass, 
$m_{H^{+}}$ (in GeV). Magenta points have been subject to no cuts, 
blue points comply with the bounds on the masses (LEP+LHC), red points
satisfy all bounds except BR($B_{u}\to\tau\nu$) and green points
satisfy all bounds. Right panel: BR($B_{u}\to\tau\nu$) versus $m_{H^{+}}$.
Red points satisfy only the bounds on the masses (LEP+LHC) while green points
comply with all bounds.}
\label{fig:NUHM:deltar} 
\end{figure*}
As can be verified from the left-hand panel of
Fig.~\ref{fig:NUHM:deltar}, one could in principle have
semi-constrained regimes leading to sizable values of $R_{K}$,
$\mathcal{O}(10^{-2})$.  Once all (collider and low-energy) bounds
have been imposed, one has at most $\Delta r\lesssim10^{-4}$ (in
association with $m_{H^+}\gtrsim500$ GeV). Moreover, it is interesting
to notice that SUSY contributions to BR($B_{u}\to\tau\nu$), which
become non-negligible for lighter $H^{\pm}$, have a negative
interference with those of the SM, lowering the latter BR to values
below the current experimental bound. This can be seen on the
right-hand panel of Fig.~\ref{fig:NUHM:deltar}. We will return to this
topic in greater detail in the following subsection, when addressing
the unconstrained MSSM.

\subsection{Unconstrained MSSM}

To conclude the numerical discussion, and to allow for a better comparison
between our approach and those usually followed in other recent analyses
(for instance~\cite{Masiero:2008cb,Girrbach:2012km}), we conduct
a final study of the unconstrained, low-energy MSSM. Thus, and in
what follows, we make no hypothesis concerning the source of lepton
flavour violation, nor on the underlying mechanism of SUSY breaking.
Massive neutrinos are introduced by hand (no assumption being made
on their nature), and although charged interactions do violate lepton
flavour, as parametrised by the $U_{\text{MNS}}$ matrix, no sizable
contributions to BR($\mu\to e\gamma$) should be expected, as these
would be suppressed by the light neutrino masses. At low-energies,
no constraints (other than the relevant experimental bounds) are imposed
on the SUSY spectrum (for simplicity, we will assume a common value
for all sfermion trilinear couplings at the low-scale, $A_i=A_0$). 
The soft-breaking slepton masses
are allowed to be non-diagonal, so that a priori a non-negligible
mixing in the slepton sector can occur. In order to better correlate
the source of flavour violation at the origin of $\Delta r$ with
the different experimental bounds, we again allow for a single FV entry
in the slepton mass matrices: $\delta_{31}^{RR}\sim0.5$ (otherwise
setting all other $\delta_{ij}^{XY}=0$). 

\begin{figure*}[htb]
\centering
\begin{tabular}{cc}
\includegraphics[clip, width=0.49\linewidth]{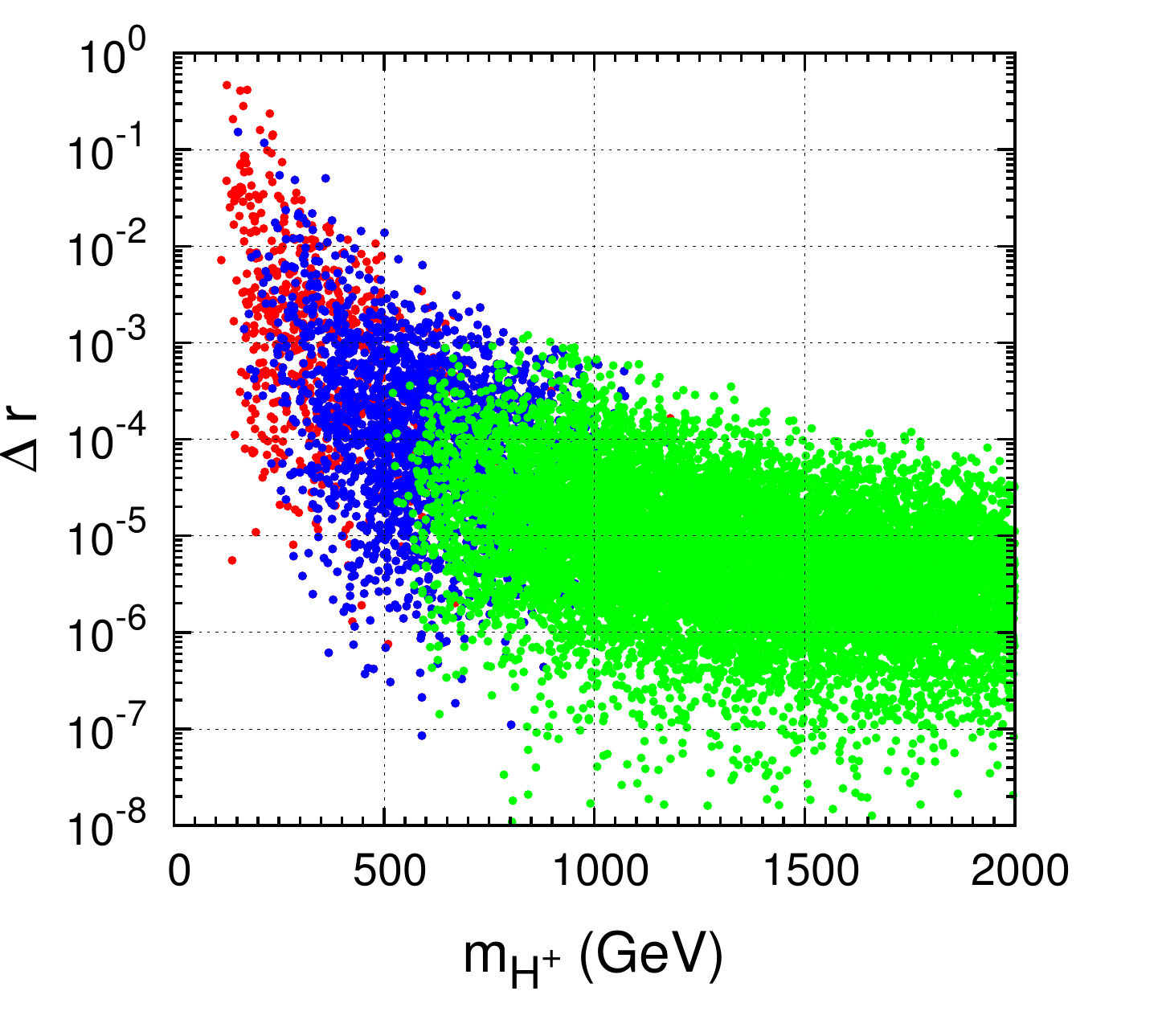}
&\includegraphics[clip, width=0.49\linewidth]{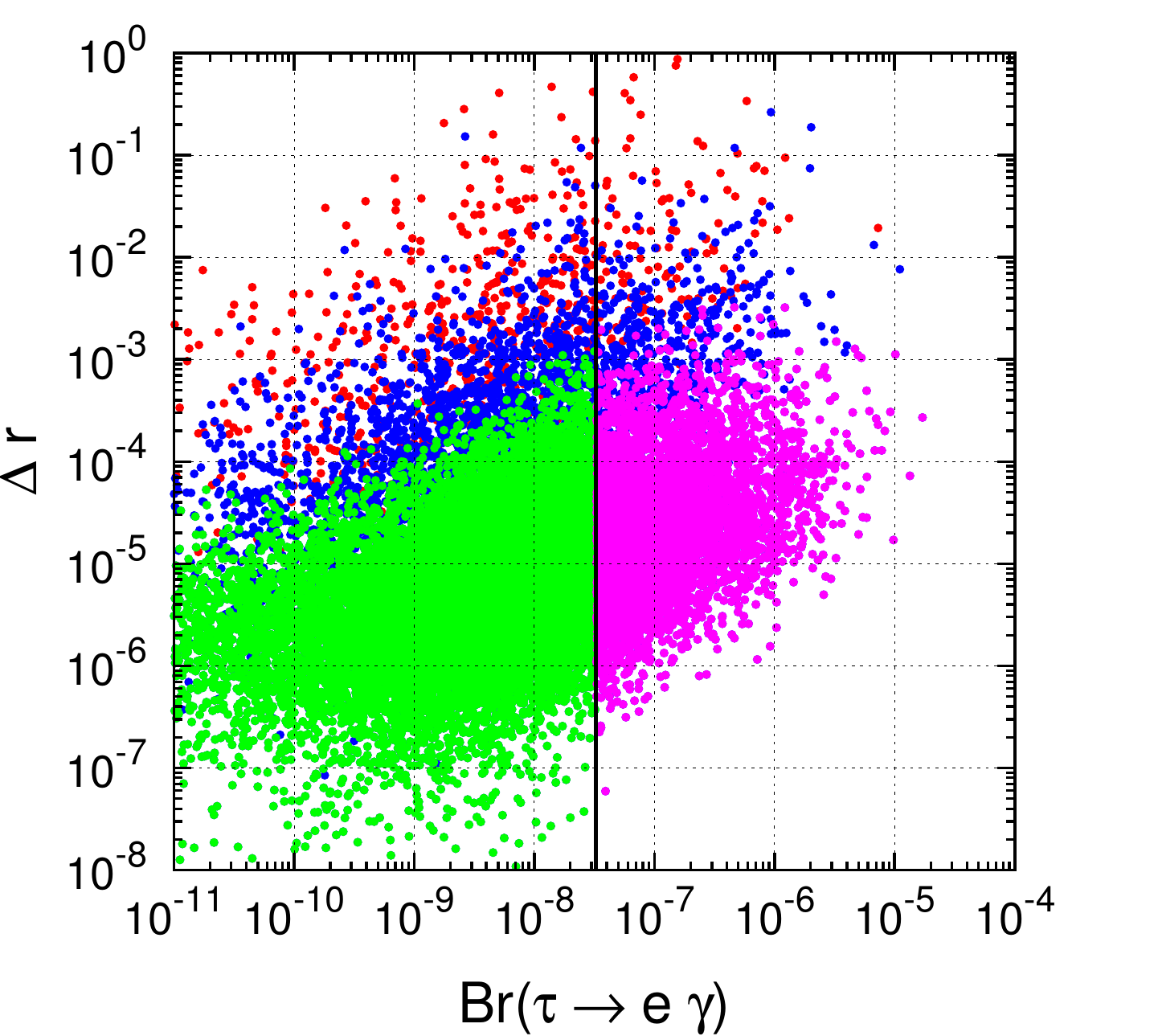}
\end{tabular}
\caption{Ranges of variation of $\Delta r$ in the unconstrained MSSM
as a function of $m_{H^+}$ (left panel), and
as a function of BR($\tau\to e\gamma$) (right panel). The different
input parameters were varied as in Table~\ref{tab:ranges} (notice that
$\delta_{31}^{RR}=0.5$). On the left panel
red points satisfy the bounds on the
masses (LEP+LHC), blue points satisfy all bounds except
$\textrm{BR}\left(B_{u}\to\tau\nu\right)$ and green points comply with all
bounds. Similar colour code on the right panel, except that  
blue points now comply with all bounds except
$\textrm{BR}\left(B_{u}\to\tau\nu\right)$ and BR($\tau\to e\gamma$)
while magenta denotes points only failing the bound on BR($\tau\to e\gamma$).}
\label{fig:DeltaRComparisonCuts} 
\end{figure*}

\begin{table*}[htb]
\begin{center}
\begin{tabular}{cccccccccccc}
\hline \\[-2mm]
 & $\mu$  & $m_{A}$  & $M_{1}$, $M_{2}$  & $M_{3}$  & $A_{0}$  &
 $m_{L}$  & $m_{R}$  & $m_{Q},m_{U},m_{D}$  & $\tan\beta$  &
 $\delta_{31}^{RR}$  & other $\delta_{ij}^{XY}$ \\[+1mm]
\hline\\[-3mm]
Min  & 100  & 50 & 100  & 1100  & -1000  & 100  & 100  & 1200  & 30 &
0.5 & 0\\
Max  & 3000  & 1500  & 2500  & 2500 & 1000  & 2200 & 2500 & 5000  & 60
& 0.5 & 0\\[+1mm]
\hline
\end{tabular}
\end{center}
\caption{Range of variation of the unconstrained MSSM
parameters (dimensionful parameters in GeVs). 
$A_{0}$ denotes the common value of the low-energy
 sfermion trilinear couplings.
}
\label{tab:ranges} 
\end{table*}
In our scan we have varied the input parameters in the ranges collected
in Table~\ref{tab:ranges}. We have also applied all relevant constraints
on the low-energy observables,
Eq.~(\ref{eq:cLFVbounds1}-\ref{eq:Bbounds2}), 
as well as the constraints on the SUSY
spectrum~\cite{PDG,LHC:2011}. 
In particular we have assumed the conservative limits 
\begin{equation}
m_{\tilde{q}_{L,R}}>1000\,\text{GeV}\,,\quad\quad m_{\tilde{g}}>1000\,\text{GeV}\,.
\end{equation}
Concerning the light Higgs boson mass, no constraint was explicitly 
imposed. We just notice here that values close to 
125 GeV~\cite{LHC:Higgs:2012}, or even
larger, are easily achievable due to the heavy squark masses. 
\begin{figure*}[htb]
\centering
\begin{tabular}{cc}
\includegraphics[clip,width=0.49\linewidth]{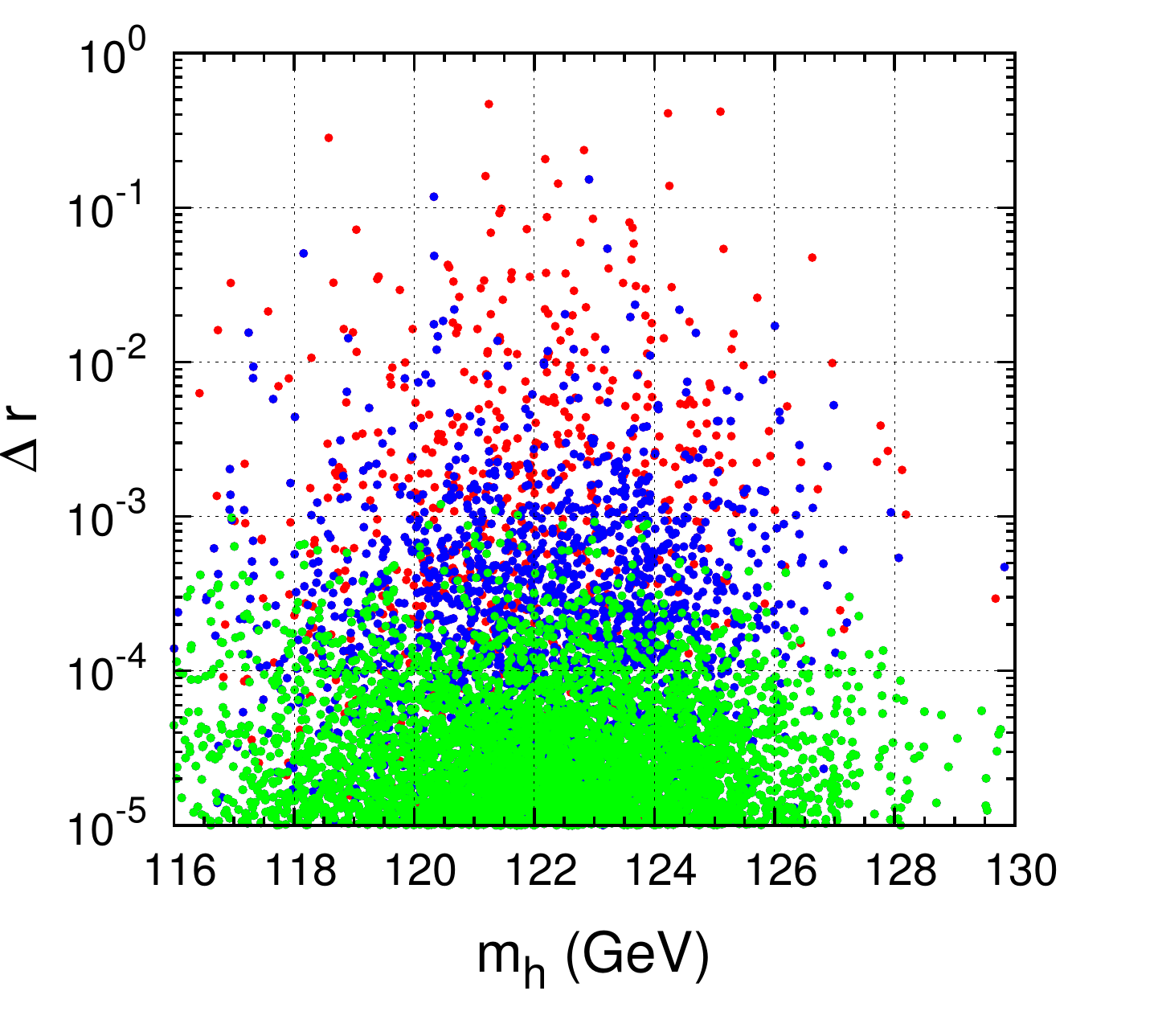}
&\includegraphics[clip,width=0.49\linewidth]{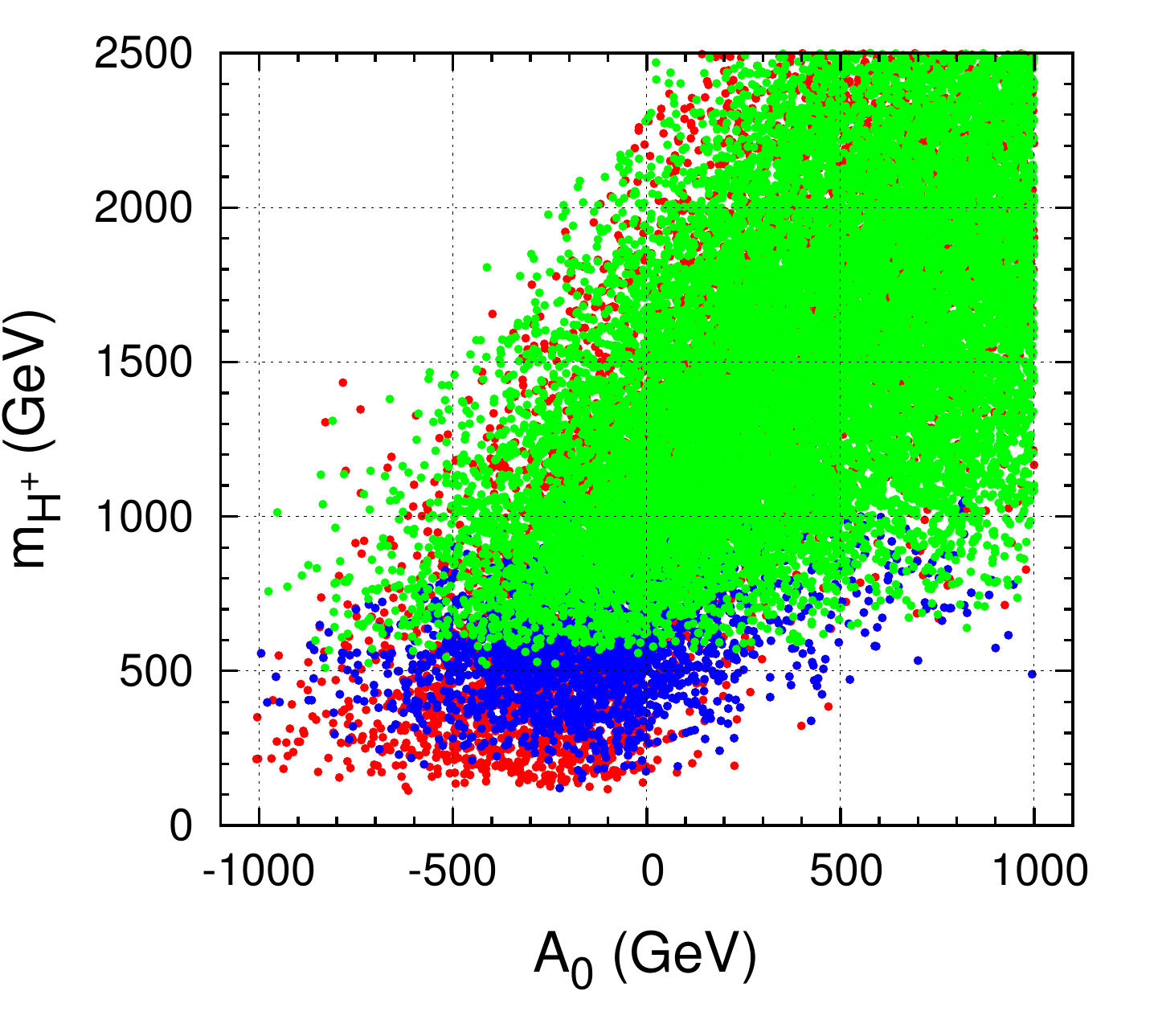}
\end{tabular}
\caption{Left panel: $\Delta r$ as a function of the lightest Higgs
boson mass $m_{h}$ (in GeV) for the range of parameters shown in
Table~\ref{tab:ranges}. Red points satisfy the bounds on the
spectrum (LEP+LHC), blue points satisfy all bounds except
$\textrm{BR}\left(B_{u}\to\tau\nu\right)$ and green points satisfy all
bounds. Right panel: $m_{H^{+}}$ versus $A_{0}$ 
(both in GeV), with the same colour code.
Leading to both plots, the different
input parameters were varied as in Table~\ref{tab:ranges}.}
\label{fig:DeltaR-mh0}
\end{figure*}
This can be observed from the left panel of Fig.~\ref{fig:DeltaR-mh0},
where we display the output of the above scan, presenting the values
of $\Delta r$ versus the associated light Higgs boson mass, $m_{h}$. 
As expected, no explicit correlation between $m_{h}$ and $\Delta r$ is
manifest, nor with the other (relevant) flavour-related low-energy
bounds. For completeness, and to better illustrate the following
discussion, we present on the right-hand panel of
Fig.~\ref{fig:DeltaR-mh0} the charged Higgs mass as a function of 
$A_0$, again under a
colour scheme denoting the experimental bounds applied in each case.
Identical to what was observed in Fig.~\ref{fig:NUHM:deltar} (notice
that NUHM models correspond, at low-energies, to a subset of these
general cases), regimes of very light charged Higgs are indeed
present, in association with small to moderate 
(negative) regimes for $A_0$. 
Nevertheless, these regimes
 - which could potentially enhance $\Delta r$ - are likewise excluded
 due to a strong conflict with $\textrm{BR}\left(B_{u}\to\tau\nu\right)$.
This can be further confirmed from the left panel of
Fig.~\ref{fig:DeltaRComparisonCuts}, where we display the possible
range of variation for $\Delta r$ as a function of $m_{H^+}$,
colour-coding the different applied bounds.

As can be seen from both panels of Fig.~\ref{fig:DeltaRComparisonCuts},
values $\Delta r\approx\mathcal{O}(10^{-2},10^{-1})$ could be obtainable, in
agreement with
Refs~\cite{Girrbach:2012km,Masiero:2005wr,Masiero:2008cb,Ellis:2008st}.
However, the situation is substantially altered when one takes into
account the current experimental bounds on $B$ decays 
($B_{u}\to\tau\nu$ and $B_{s}\to\mu^{+}\mu^{-}$) and $\tau\to e\gamma$.
As is manifest from the left panel of Fig.~\ref{fig:DeltaRComparisonCuts},
once experimental bounds - other than $B_{u}\to\tau\nu$ - are
imposed, one could in principle have $\Delta
r^{\text{max}}\approx\mathcal{O}(10^{-2})$; however, taking into account 
the limits from BR($B_{u}\to\tau\nu$), one is now led to 
$\Delta r\lesssim 10^{-3}$.

A few comments are in order regarding the impact of the different
low-energy bounds from radiative tau decays and $B$-physics
observables.
Firstly, let us consider the $\tau\to e\gamma$ decay: although
directly depending on 
$\delta_{31}^{RR}$, its amplitude is (roughly) suppressed by the 
fourth power of the average SUSY scale, $m_{\text{SUSY}}$. 
As can be seen from Eqs.~(\ref{eq:deltar:approx:X}, \ref{eq:deltar:approx}),
$\Delta r$ only depends on the charged Higgs mass - if the latter
is assumed to be an EW scale parameter, $\Delta r$
will be thus independent of $m_{\text{SUSY}}$ in these unconstrained models.
As such, it is possible to evade the $\tau\to e\gamma$ bound by 
increasing the soft SUSY masses, and this can indeed be seen from the right-hand
panel of Fig.~\ref{fig:DeltaRComparisonCuts}, where a number of
``blue'' points are found to lie below the BR($\tau\to e\gamma$) bound.

\noindent
Secondly, the $B_{s}\to\mu^{+}\mu^{-}$ decay 
can be a severe constraint regarding the SUSY contributions to
$\Delta r$ in the case of constrained models
(see, e.g., Figs.~\ref{fig:cMSSM:seesaw} and \ref{fig:cMSSM:survey}).
We notice that $B_{s}\to\mu^{+}\mu^{-}$ is approximately
proportional to $A_{0}^{2}$ (see for instance \cite{Isidori:2006pk})
while $\Delta r$ shows no such dependence: thus a regime of small 
trilinear couplings easily allows evade the $B_{s}\to\mu^{+}\mu^{-}$
bounds.

\noindent
Finally, let us discuss the $B_{u}\to\tau\nu$ bounds. Notice that 
this is a process essentially identical to the charged kaon decays at
the origin of the
$R_{K}$ ratio (the only difference being
that the $K^{+}$ meson is to be replaced by a $B_{u}$ and the
$e$/$\mu$ in the decay products by a kinematically allowed $\tau$),  
and hence its tree-level decay width can be inferred 
from Eqs.~(\ref{eq:SM:Pdecays}) and
(\ref{eq:kaon:gamma:smsusy}).
Due to a negative
interference between the SM and the MSSM contributions, given by the term
proportional to $\tan^2\beta/m^2_{H^\pm}$ in
Eq.~(\ref{eq:kaon:gamma:smsusy}), regimes of low  $m_{H^+}$ 
lead to excessively small values of $B_{u}\to\tau\nu$ (below the
experimental bound), effectively setting a lower bound for 
for $m^2_{H^\pm}$ (see right panel of Fig.~\ref{fig:NUHM:deltar}, in
relation to the discussion of NUHM models). In turn, this excludes 
regimes of $m_{H^+}$ associated to sizable values of $\Delta r$, as
is clear from the comparison of the ``blue'' and ``green'' regions of the 
left panel of Fig.~\ref{fig:DeltaRComparisonCuts}.

\bigskip
In summary, we conclude that saturating the experimental bound on
$R_{K}$ clearly proves to be extremely difficult (if not impossible),
even in the unconstrained MSSM, especially in view of the stringent
constraints from $B_{u}\to\tau\nu$.

\section{Conclusions}
\label{sec:concs}

In this work we have revisited supersymmetric contributions to 
$R_{K}=\Gamma\left(K\rightarrow e\nu\right)$$/\Gamma\left(K\rightarrow\mu\nu\right)$,
considering the potential of a broad class of constrained SUSY models
to saturate the current measurement of $R_{K}$. We based our analysis
in a full computation of the one-loop corrections to the $\nu\ell H^{+}$
vertex; we have also derived (when possible) illustrative analytical
approximations, which in addition to offering a more transparent understanding
of the r\^ole of the different parameters, also allow to establish a
bridge between our results and previous ones in the literature. 
Our analysis further revisited the $R_{K}$
observable in the
light of new experimental data, arising from flavour physics as
well as from collider searches.

We numerically evaluated the contributions to $R_{K}$ arising in
the context of different minimal supergravity inspired models which
account for observed neutrino data, further considering the possibility
of accommodating a near future observation of a $\mu\to e\gamma$
decay. As expected from the (mostly) $LL$ nature of the radiatively
induced charged lepton flavour violation, type I and II seesaw mechanisms
implemented in the cMSSM provide minimal contributions to $R_{K}$,
thus implying that such cMSSM SUSY seesaws cannot saturate the present
value for $\Delta r$.

We then considered unified models where the
flavour-conserving hypothesis on the $RR$ slepton sector is relaxed by
allowing a non-vanishing $\delta_{31}^{RR}$ ($e-\tau$ sector). 
In all
models, special attention was given to experimental
constraints, especially four observables which turn out to play a
particularly relevant r\^ole: the recent interval for 
the lightest neutral Higgs
boson mass provided by the CMS and ATLAS collaborations,
BR($B_{s}\to\mu^{+}\mu^{-}$), BR($B_{u}\to\tau\nu$) and BR($\tau\to
e\gamma$). These last two 
exhibit a dependence on $m_{H^+}$ ($B_{u}\to\tau\nu$) and on
$\delta_{31}^{RR}$ ($\tau\to e\gamma$)  similar to that of $\Delta r$.
The SUSY contributions to $\Delta r$ are thus maximised in a regime in
which $m_{H^+}$ and
$\delta_{31}^{RR}$ are such that the experimental limits for
$B_{u}\to\tau\nu$ and $\tau\to e\gamma$ are simultaneously saturated;
in this regime one must then accommodate 
the bounds on other observables, such as $m_h$ and
BR($B_{s}\to\mu^{+}\mu^{-}$).
For a minimal deviation from a pure cMSSM scenario
allowing for non-vanishing values of $\delta_{31}^{RR}$,  
we can have  values for $\Delta r$ at most of the order of $10^{-6}$.
In fact, the requirement of having a Higgs boson 
mass of 125-126 GeV is much more constraining on the cMSSM
parameter space than, for instance $B_{s}\to\mu^{+}\mu^{-}$
(which is sub-dominant, and can be overcome by variations of the
trilinear coupling, $A_0$).
In order to have $\Delta r \sim \mathcal{O}(10^{-6})$, one must 
significantly increase $\delta_{31}^{RR}$ so to marginally overlap 
the regions of $m_h \sim 125$ GeV, while still in agreement with 
$\tau\to e\gamma$.

Models where the charged Higgs mass can be 
significantly lowered,
as is the case of NUHM models, allow to increase the SUSY
contributions to $\Delta r$, which can be as large as $10^{-4}$
(larger values being precluded due to $B_{u}\to\tau\nu$ decay constraints).

More general models, as the unconstrained MSSM
realised at low-energies, offer more degrees of freedom, and the
possibility to better accommodate/evade the different experimental 
constraints. 
In the  unconstrained MSSM, one can find values of $\Delta r$
one order of magnitude larger, $\Delta r \sim \mathcal{O}(10^{-3})$.
Again, any further augmentation is precluded due to incompatibility
with the bounds on $B_{u}\to\tau\nu$.

However $\Delta r \sim \mathcal{O}(10^{-3})$ still remains one order of
magnitude shy of the current experimental sensitivity to $R_{K}$,
and also substantially lower than some of the values previously found
in the literature. As such, if SUSY is indeed discovered, and unless there
is significant progress in the experimental sensitivity to $R_{K}$, it
seems unlikely that the contributions to $R_{K}$ of the SUSY models studied here
will be testable in the near future. On the other hand, any
near-future measurement of $\Delta r$ larger than
$\mathcal{O}(10^{-3})$ would unambiguously point towards a scenario
different than those here addressed (mSUGRA-like seesaw, NUHM and the
phenomenological MSSM).

It should be kept in mind that the analysis presented here focused
on the impact of LFV interactions. Should the discrepancy between
the SM and experimental observations turn out to be much smaller than
$10^{-4}$, a more detailed approach and evaluation will then be necessary.

\section*{Acknowledgments}

R.M.F. is thankful for the hospitality of the LPC Clermont-Ferrand.
The work of R.M.F has been supported by \textit{Funda\-\c{c}\~ao para a Ci\^encia
e a Tecnologia} through the fellowship SFRH/BD/47795/2008. R.M.F.
and J. C. R. also acknowledge the financial support from the EU Network
grant UNILHC PITN-GA-2009-237920 and from \textit{Funda\c{c}\~ao para a
Ci\^encia e a Tecnologia} grants CFTP-FCT UNIT 777, CERN/FP/83503/2008
and PTDC/FIS/102120/2008. A. M. T. acknowledges partial support from the
European Union FP7 ITN-INVISIBLES (Marie Curie Actions, PITN-
GA-2011-289442).

\appendix

\section{Renormalisation of the $\nu\ell H^{+}$ vertex}

\label{sec:app1}
In what follows we detail the computation leading to 
Eqs. (\ref{eq:epsilondelta}-\ref{eq:delta:def}),
and we further refer to~\cite{Bellazzini:2010gn} for a similar analysis.
As expected, loop effects contribute to both kinetic and mass terms
of charged leptons as well as to the ${\nu}\ell H^{+}$ vertex: 
\begin{align}
\mathcal{L}_{0}^{H^{\pm}}\!\!  = &
i\,\overline{\ell}_{L}
\left(\mathbf{1}+
  \eta_{L}^{\ell}\right)\slashed{\partial}\ell_{L}+i\,\overline{\ell}_{R} 
\left(\mathbf{1}+\eta_{R}^{\ell}\right)
\slashed{\partial}\ell_{R}\nonumber\\[+1mm]
&\!\!  +i\,
\overline{\nu}_{L}\left(\mathbf{1}+\eta_{L}^{\nu}\right)  
\slashed{\partial}\nu_{L}-
\left[\overline{\ell}_{L}
  \left(M^{l0}+\eta_{m}^{\ell}\right)\ell_{R}+\textrm{h.c.}\right]\nonumber   
\\[+1mm] 
  &\!\!  +\left[\overline{\nu}_{L}\left(2^{3/4}G_{F}^{1/2}\,\tan\beta\,
     M^{l0}+\eta^{H}\right)\ell_{R}H^{+}+\textrm{h.c.}\right]\,.
\end{align}
 Here $M^{l0}$ denotes the bare charged lepton mass and the $\eta$'s
correspond to loop contributions to the various terms. The (new) kinetic
terms can be recast into a canonical form by means of unitary rotations
of the fields ($K_{L}^{\ell}$, $K_{R}^{\ell}$, $K_{L}^{\nu}$),
which are then renormalised by diagonal transformations ($\hat{Z}_{L}^{\ell}$,
$\hat{Z}_{R}^{\ell}$, $\hat{Z}_{L}^{\nu}$): 
\begin{align}
\ell_{L}^{\text{old}} \! = &
K_{L}^{\ell}\left(\hat{Z}_{L}^{\ell}\right)^{-\frac{1}{2}}\!\! \ell_{L}^{\text{new}}
\ \textrm{;} \quad
\hat{Z}_{L}^{\ell}={K_{L}^{\ell}}^{\dagger}
\left(\mathbf{1}+\eta_{L}^{\ell} \right)K_{L}^{\ell}\,,\\  
\ell_{R}^{\text{old}} \! = & K_{R}^{\ell}
\left(\hat{Z}_{R}^{\ell}\right)^{-\frac{1}{2}}\!\! \ell_{R}^{\text{new}} 
\ \textrm{;}\quad \hat{Z}_{R}^{\ell}={K_{R}^{\ell}}^{\dagger}
\left(\mathbf{1}+\eta_{R}^{\ell}\right)K_{R}^{\ell}\,,\\ 
\nu_{L}^{\text{old}} \! = & K_{L}^{\nu}\left(\hat{Z}_{L}^{\nu}
\right)^{-\frac{1}{2}}\!\! \nu_{L}^{\text{new}} \ \textrm{;}\quad
\hat{Z}_{L}^{\nu}={K_{L}^{\nu}}^{\dagger}
\left(\mathbf{1}+\eta_{L}^{\nu}\right)K_{L}^{\nu}\,. 
\end{align}
 Two unitary rotation matrices ($R_{L}^{\ell}$, $R_{R}^{\ell}$)
are further required to diagonalise the charged lepton mass matrix,
and one finally has 
\begin{eqnarray}
\ell_{L}^{\text{old}} & = & K_{L}^{\ell}\left(\hat{Z}_{L}^{\ell}\right)^{-\frac{1}{2}}\, R_{L}^{\ell}\,\ell_{L}^{\text{new}}\,,\\
\ell_{R}^{\text{old}} & = & K_{R}^{\ell}\left(\hat{Z}_{R}^{\ell}\right)^{-\frac{1}{2}}\, R_{R}^{\ell}\,\ell_{R}^{\text{new}}\,,\\
\nu_{L}^{\text{old}} & = & K_{L}^{\nu}\left(\hat{Z}_{L}^{\nu}\right)^{-\frac{1}{2}}\, R_{L}^{\ell}\,\nu_{L}^{\text{new}}\,.
\end{eqnarray}
 In the new basis, the mass terms now read 
\begin{align}
\mathcal{L}^{\textrm{mass}}\equiv&-\overline{\ell}_{L}\,
M^{l}\,\ell_{R}
+\textrm{h.c.}
\nonumber \\
=&-\overline{\ell}_{L}\,{R_{L}^{\ell}}^{\dagger}
\left[\left(\hat{Z}_{L}^{\ell}\right)^{-\frac{1}{2}}
  \,{K_{L}^{\ell}}^{\dagger}\left(M^{l0} 
+\eta_{m}^{\ell}\right) \right. 
\nonumber\\
&\left.\hskip 17mm
 K_{R}^{\ell}\left(\hat{Z}_{R}^{l}
  \right)^{-\frac{1}{2}}\right]\,
R_{R}^{\ell}\,\ell_{R}+\textrm{h.c.}\,. 
\end{align}
 The above equation relates the unknown parameter $M^{l0}$ with the
physical mass matrix $M^{l}$. Using the latter to rewrite the ${\nu}\ell H^{+}$
vertex one finds 
\begin{eqnarray}
\mathcal{L}^{H^{\pm}} & \equiv & \overline{\nu}_{L}\,
Z^{H}\,\ell_{R}\, H^{+}+\textrm{h.c.}\,, 
\end{eqnarray}
 where 
\begin{align}
Z^{H}  = &
2^{3/4}G_{F}^{1/2}\,\tan\beta\,{R_{L}^{\ell}}^{\dagger}
\left(\hat{Z}_{L}^{\nu}\right)^{-\frac{1}{2}}{K_{L}^{\nu}}^{\dagger}\, 
 K_{L}^{\ell}\left(\hat{Z}_{L}^{\ell}\right)^{\frac{1}{2}}\, R_{L}^{\ell}
\, M^{l}\nonumber \\
   & +{R_{L}^{\ell}}^{\dagger}\left(\hat{Z}_{L}^{\nu}
\right)^{-\frac{1}{2}}\,{K_{L}^{\nu}}^{\dagger}
\left(-2^{3/4}G_{F}^{1/2}\,\tan\beta\,\eta_{m}^{\ell}
+\eta^{H}\right)\nonumber\\
& \hskip 5mm K_{R}^{\ell}
\left(\hat{Z}_{R}^{\ell}\right)^{-\frac{1}{2}}\, K_{R}^{\ell}\,.
\end{align}
 To one-loop order, this exact expression simplifies to 
\begin{align}
Z^{H} = & 2^{3/4}G_{F}^{1/2}\,\tan\beta\,\left[\left(\mathbf{1}+\frac{\eta_{L}^{\ell}}{2}-\frac{\eta_{L}^{\nu}}{2}\right)\, M^{l}-\eta_{m}^{\ell}\right]+\eta^{H}\,.
\end{align}
 The expressions for the $\eta$'s can be computed from the relevant
Feynman diagrams (assuming zero external momenta): 
\begin{align}
-\left(4\pi\right)^{2}\hskip -1mm &\left(\eta_{m}^{\ell}\right)_{ij}=  \,
N_{i\alpha\beta}^{R\left(\ell\right)}N_{j\alpha\beta}^{L\left(\ell\right)*}
m_{\chi_{\alpha}^{0}} 
B_{0}\left(0,m_{\chi_{\alpha}^{0}}^{2},m_{\widetilde{\ell}_{\beta}}^{2}\right) 
\nonumber\\
+ & C_{i\alpha\beta}^{R\left(\ell\right)}
C_{j\alpha\beta}^{L\left(\ell\right)*}m_{\chi_{\alpha}^{\pm}}  
B_{0}\!\left(0,m_{\chi_{\alpha}^{\pm}}^{2},m_{\widetilde{\nu}_{\beta}}^{2}\right)\!,
\label{eq:app:etaellm}\\[+1mm]
-\left(4\pi\right)^{2}\hskip -1mm &\left(\eta_{R}^{\ell}\right)_{ij}= 
 \,
N_{i\alpha\beta}^{L\left(\ell\right)}N_{j\alpha\beta}^{L\left(\ell\right)*}
B_{1}\left(0,m_{\chi_{\alpha}^{0}}^{2},m_{\widetilde{\ell}_{\beta}}^{2}\right)
\nonumber\\
+& C_{i\alpha\beta}^{L\left(\ell\right)}C_{j\alpha\beta}^{L\left(\ell\right)*}
B_{1}\left(0,m_{\chi_{\alpha}^{\pm}}^{2},m_{\widetilde{\nu}_{\beta}}^{2}\right),
\label{eq:app:etaellR}\\[+1mm]
-\left(4\pi\right)^{2}\hskip -1mm &\left(\eta_{L}^{\ell}\right)_{ij}= 
 \, N_{i\alpha\beta}^{R\left(\ell\right)}N_{j\alpha\beta}^{R\left(\ell\right)*}
B_{1}\left(0,m_{\chi_{\alpha}^{0}}^{2},m_{\widetilde{\ell}_{\beta}}^{2}\right)
\nonumber\\
+ & C_{i\alpha\beta}^{R\left(\ell\right)}C_{j\alpha\beta}^{R\left(\ell\right)*}
B_{1}\left(0,m_{\chi_{\alpha}^{\pm}}^{2},m_{\widetilde{\nu}_{\beta}}^{2}
\right)\!,\label{eq:app:etaellL}\\[+1mm] 
-\left(4\pi\right)^{2}\hskip -1mm &\left(\eta_{L}^{\nu}\right)_{ij}= 
 \,
N_{i\alpha\beta}^{R\left(\nu\right)}N_{j\alpha\beta}^{R\left(\nu\right)*}
B_{1}\left(0,m_{\chi_{\alpha}^{0}}^{2},m_{\widetilde{\nu}_{\beta}}^{2}\right)
\nonumber\\
+ & C_{i\alpha\beta}^{R\left(\nu\right)}C_{j\alpha\beta}^{R\left(\nu\right)*}
B_{1}\left(0,m_{\chi_{\alpha}^{\pm}}^{2},m_{\widetilde{\ell}_{\beta}}^{2}\right),
\label{eq:app:etanuL}\\[+1mm]
-\left(4\pi\right)^{2}\hskip -1mm &\left(\eta^{H}\right)_{ij}= \,
C_{i\beta\gamma}^{R\left(\nu\right)}N_{j\alpha\gamma}^{L\left(\ell\right)*}
\left[D_{\beta\alpha2}^{L\left(S^{+}\right)*}m_{\chi_{\alpha}^{0}}
m_{\chi_{\beta}^{\pm}}\right.\nonumber\\
& \left.\hskip 25mm
C_{0}\left(0,0,0,m_{\chi_{\alpha}^{0}}^{2},m_{\chi_{\beta}^{\pm}}^{2},m_{\widetilde{\ell}_{\gamma}}^{2}\right)\right.\nonumber \\
+ & \left. D_{\beta\alpha2}^{R\left(S^{+}\right)*}
dC_{00}\left(0,0,0,m_{\chi_{\alpha}^{0}}^{2},m_{\chi_{\beta}^{\pm}}^{2},m_{\widetilde{\ell}_{\gamma}}^{2} \right)\right]\nonumber \\
 & +\, N_{i\alpha\gamma}^{R\left(\nu\right)}
 C_{j\beta\gamma}^{L\left(\ell\right)*}\left[D_{\beta\alpha2}^{L\left(S^{+}\right)*}
   m_{\chi_{\alpha}^{0}}m_{\chi_{\beta}^{\pm}} \right.\nonumber\\
&\left.\hskip 30mm
C_{0}\left(0,0,0,m_{\chi_{\alpha}^{0}}^{2},m_{\chi_{\beta}^{\pm}}^{2},m_{\widetilde{\nu}_{\gamma}}^{2}\right)\right.\nonumber \\
 & \left.+\, D_{\beta\alpha2}^{R\left(S^{+}\right)*}
dC_{00}\left(0,0,0,m_{\chi_{\alpha}^{0}}^{2},m_{\chi_{\beta}^{\pm}}^{2},m_{\widetilde{\nu}_{\gamma}}^{2}\right)\right]\nonumber \\
 & +\,
 N_{i\alpha\beta}^{R\left(\nu\right)}
N_{j\alpha\gamma}^{L\left(\ell\right)*}
g_{2\gamma\beta}^{\left(S^{+}\widetilde{\ell}\widetilde{\nu}^{*}\right)}
m_{\chi_{\gamma}^{0}}\nonumber\\
&\hskip 20mm
C_{0}\left(0,0,0,m_{\widetilde{\ell}_{\gamma}}^{2},m_{\widetilde{\nu}_{\beta}}^{2},m_{\chi_{\alpha}^{0}}^{2}\right)\,,\label{eq:app:etaH}
\end{align}
with $B_{0,1},C_{0},C_{0,0}$ denoting the usual loop integral functions
\begin{align}
 & B_{0}\left(0,x,y\right)=\,\Delta_{\varepsilon}+1
 -\frac{x\log\frac{x}{\mu^{2}}-y\log\frac{y}{\mu^{2}}}{x-y}\,,\\
 & B_{1}\left(0,x,y\right)=\,-\frac{1}{2}\left[\Delta_{\varepsilon}
+\frac{3x-y}{2\left(x-y\right)}\right.\nonumber\\
&\left.\hskip 25mm
-\log\frac{y}{\mu^{2}}+\left(\frac{x}{x-y}\right)^{2}\log\frac{y}{x}\right]\,,\label{eq:app:B1}\\
 & C_{0}\left(0,0,0,x,y,z\right)=\,\frac{xy\log\frac{x}{y}+yz\log\frac{y}{z}+zx\log\frac{z}{x}}{\left(x-y\right)\left(y-z\right)\left(z-x\right)}\,,\\
 & dC_{00}\left(0,0,0,x,y,z\right)=\,\Delta_{\varepsilon}+1\nonumber\\
&\hskip -2mm
+\frac{x^{2}\left(y-z\right)\log\frac{x}{\mu^{2}}+y^{2}\left(z-x\right)\log\frac{y}{\mu^{2}}+z^{2}\left(x-y\right)\log\frac{z}{\mu^{2}}}{\left(x-y\right)\left(y-z\right)\left(z-x\right)}\,.
\end{align}
 Here $d=4-\varepsilon$, $\mu$ is the regularisation parameter and
$\Delta_{\varepsilon}=\frac{2}{\varepsilon}-\gamma+\log4\pi$. For
the couplings notation we followed~\cite{romao:MSSM}.

The comparison of the above expressions with the corresponding ones
derived in Ref.~\cite{Bellazzini:2010gn}, reveals a fair agreement;
we nevertheless notice that the neutralino and chargino masses are
absent from the analogous of Eq.~(\ref{eq:app:etaellm}), and that
the order of the arguments of $B_{1}$ in Eqs.~(\ref{eq:app:etaellR},
\ref{eq:app:etaellL}, \ref{eq:app:etanuL}) appears reversed. Moreover,
we find small discrepancies (which cannot be accounted by the distinct
notations used) in the expressions for $\eta_{m}^{\ell}$ and $\eta_{H}$,
cf. Eq.~(\ref{eq:app:etaellm}) and Eq.~(\ref{eq:app:etaH}), respectively.

\section{SUSY seesaw models}

\label{app:seesaw} In its different realisations, the seesaw mechanism
offers one of the most appealing explanations for the smallness of
neutrino masses and the pattern of neutrino mixing angles. Moreover,
when embedded in the framework of SUSY models - the so-called SUSY
seesaw - the seesaw offers the interesting feature that flavour violation
in the neutrino sector (encoded in non-diagonal neutrino Yukawa couplings)
can radiatively induce flavour violation in the slepton sector at
low-energies~\cite{Borzumati:1986qx}, leading to potentially sizable
contributions to a large array of observables.

In what follows we briefly summarise the most relevant features of
different realisations of the seesaw mechanism. In particular, we
will consider {}``high-scale'' seesaws, i.e., where the additional
states are assumed to be much heavier than the electroweak scale (in
association with large values of the corresponding couplings).

\subsection{Type I SUSY seesaw}

\label{app:seesawI}

In a type I SUSY seesaw, the MSSM superfield content is extended by
three right-handed Majorana neutrino superfields. The lepton superpotential
is thus extended as 
\begin{equation}
\mathcal{W}_{\text{I}}^{\text{lepton}}=\hat{N}^{c} Y^{\nu} \hat{L}
\hat{H}_{2}+\hat{E}^{c} Y^{l} \hat{L} \hat{H}_{1} + \frac{1}{2}
\hat{N}^{c}  M_{N} \hat{N}^{c} ,\label{eq:WleptonI:def}
\end{equation}
 where, and without loss of generality, one can work in a basis where
both $Y^{l}$ and $M_{N}$ are diagonal 
($Y^{l}=\operatorname{diag}(Y^{e},Y^{\mu},Y^{\tau})$,
$M_{N}=\operatorname{diag}(M_{N_{1}},M_{N_{2}},M_{N_{3}})$). The
relevant slepton soft-breaking terms are now 
\begin{align}
\mathcal{V}_{\text{soft\ I}}^{\text{slepton}}=& 
m_{\tilde{L}}^{2}\,\tilde{l}_{L}\,\tilde{l}_{L}^{*}
+m_{\tilde{R}}^{2}\,\tilde{l}_{R}\,\tilde{l}_{R}^{*}
+m_{\tilde{\nu}_{R}}^{2}\,\tilde{\nu}_{R}\,\tilde{\nu}_{R}^{*}
+\!\left(A^{l}\, H_{1}\,\tilde{l}_{L}\,\tilde{l}_{R}^{*}
\right.\nonumber\\
&\left.
+A^{\nu}\, H_{2}\,\tilde{\nu}_{L}\,\tilde{\nu}_{R}^{*}
+B^{\nu}\,\tilde{\nu}_{R}\,\tilde{\nu}_{R}+\text{h.c.}\right).
\end{align}
 Should this be embedded into a cMSSM, then the additional soft breaking
parameters would also obey universality conditions at the GUT scale,
$(m_{\widetilde{\nu}_{R}})_{ij}^{2}=m_{0}^{2}$ and $(A^{\nu})_{ij}=A_{0}(Y^{\nu})_{ij}$.

In this case, the light neutrino masses are given by 
\begin{equation}
m_{\nu}^{\text{I}}\,=\,-{m_{D}^{\nu}}^{T}M_{N}^{-1}m_{D}^{\nu}\,,\label{eq:seesawI:light}
\end{equation}
 with $m_{D}^{\nu}=Y^{\nu}\, v_{2}$ ($v_{i}$ being the vacuum expectation
values (VEVs) of the neutral Higgs scalars, $v_{1(2)}=\, v\,\cos(\sin)\beta$,
with $v=174$ GeV), and where $M_{N_{i}}$ corresponds to the masses
of the heavy right-handed neutrino eigenstates. The light neutrino
matrix $m_{\nu}$ is diagonalized by the $U_{\text{MNS}}$ as $m_{\nu}^{\text{diag}}={U_{\text{MNS}}}^{T}m_{\nu}U_{\text{MNS}}$.
A convenient means of para\-metrising the neutrino Yukawa couplings,
while at the same time allowing to accommodate the experimental data,
is given by the Casas-Ibarra parametrisation~\cite{Casas:2001sr},
which reads at the seesaw scale, $M_{N}$, 
\begin{equation}
Y^{\nu}v_{2}=m_{D}^{\nu}\,=\, i\sqrt{M_{N}^{\text{diag}}}\, R\,\sqrt{m_{\nu}^{\text{diag}}}\,{U_{\text{MNS}}}^{\dagger}\,.\label{eq:seesaw:casas}
\end{equation}
 In the above, $R$ is a complex orthogonal $3\times3$ matrix that
encodes the possible mixings involving the right-handed neutrinos,
in addition to those of the low-energy sector (i.e. $U_{\text{MNS}}$)
and which can be parametrised in terms of three complex angles $\theta_{i}$
$(i=1,2,3)$. In our analysis, we assumed degenerate right-handed
neutrino masses and real parameters, so that the results are effectively
independent of the choice of the $\theta_{i}$.

Even under universality conditions at the GUT scale, the non-trivial
flavour structure of $Y^{\nu}$ will induce (through the running from
$M_{\text{GUT}}$ down to the seesaw scale, $M_{N}$) flavour mixing
in the otherwise approximately flavour conserving soft-SUSY breaking
terms. In particular, there will be radiatively induced flavour mixing
in the slepton mass matrices, manifest in the $LL$ and $LR$ blocks
of the $6\times6$ slepton mass matrix; an analytical estimation using
the leading order (LLog) approximation leads to the following corrections
to the slepton mass terms: 
\begin{align}
\label{eq:LFV:LLog}
(\Delta m_{\tilde{L}}^{2})_{_{ij}} & \,=\,-\frac{1}{8\,\pi^{2}}\,(3\, m_{0}^{2}+A_{0}^{2})\,({Y^{\nu}}^{\dagger}\, L\, Y^{\nu})_{ij}\,,\nonumber \\
(\Delta A^{l})_{_{ij}} & \,=\,-\frac{3}{16\,\pi^{2}}\, A_{0}\, Y_{ij}^{l}\,({Y^{\nu}}^{\dagger}\, L\, Y^{\nu})_{ij}\,,\nonumber \\
(\Delta m_{\tilde{R}}^{2})_{_{ij}} & \,\simeq\,0\,\,;\, L_{kl}\,\equiv\,\log\left(\frac{M_{\text{GUT}}}{M_{N_{k}}}\right)\,\delta_{kl}\,.
\end{align}
 The amount of flavour violation is encoded in the matrix elements
$({Y^{\nu}}^{\dagger}LY^{\nu})_{ij}$ of Eq.~(\ref{eq:LFV:LLog}).

\subsection{Type II SUSY seesaw}

\label{app:seesawII} The implementation of a type II SUSY seesaw
model requires the addition of at least two SU(2) triplet superfields~\cite{Rossi:2002zb}.
Should one aim at preserving gauge coupling unification, then complete
SU(5) multiplets must be added to the MSSM content. Under the SM gauge
group, the \textbf{15} decomposes as $\pmb{15}=S+T+Z$, where $S\sim(6,1,-2/3)$,
$T\sim(1,3,1)$ and $Z\sim(3,2,1/6)$. In the SU(5) broken phase (below
the GUT scale), the superpotential contains the following terms:

\begin{align}
\mathcal{W}_{\text{II}}\, 
 &=\,\frac{1}{\sqrt{2}}\,\left(Y_{T}\,\hat{L}\,\hat{T}_{1}\,
  \hat{L}+Y_{S}\,\hat{D}\,\hat{S}\,\hat{D}\right)\,
+\, Y_{Z}\,\hat{D}\,\hat{Z}\,\hat{L}\,
\nonumber\\
&
+\, Y^{d}\,\hat{D}^{c}\,\hat{Q}\,\hat{H}_{1}\,
+\, Y^{u}\,\hat{U}^{c}\,\hat{Q}\,\hat{H}_{2}\,
+\, Y^{l}\,\hat{E}^{c}\,\hat{L}\,\hat{H}_{1}\nonumber \\
\, & +\,\frac{1}{\sqrt{2}}\,\left(\lambda_{1} 
\hat{H}_{1}\,\hat{T}_{1}\,\hat{H}_{1}+\lambda_{2}\,\hat{H}_{2}\,\hat{T}_{2}\,\hat{H}_{2}\right)\,
+\, M_{T}\,\hat{T}_{1}\,\hat{T}_{2}\nonumber\\
&
+\, M_{Z}\,\hat{Z}_{1}\,\hat{Z}_{2}\,
+\, M_{S}\,\hat{S}_{1}\,\hat{S}_{2}\,+\,\mu\,\hat{H}_{1}\,\hat{H}_{2}\,,
\end{align}
 where we have omitted flavour indices for simplicity (for shortness
we will not detail the soft breaking Lagrangian here, see e.g.~\cite{Rossi:2002zb}).
After having integrated out the heavy fields, the effective neutrino
mass matrix then reads 
\begin{equation}
m_{\nu}^{\text{II}}\,=\,\frac{v_{2}^{2}}{2}\,\frac{\lambda_{2}}{M_{T}}Y_{T}\,.\label{eq:seesawII:light}
\end{equation}
 As occurs in the type I seesaw, LFV entries in the charged slepton
mass matrix are radiatively induced, and are proportional to the combination
$Y_{T}^{\dagger}Y_{T}$~\cite{Rossi:2002zb}; for example, the $LL$
block reads 
\begin{equation}
(\Delta m_{\tilde{L}}^{2})_{_{ij}}
\propto
(Y_{T}^{\dagger}\, Y_{T})_{ij} \sim \left(\frac{M_{T}}{\lambda_{2}\,
    v_{2}^{2}}\right)^{2} \left(U_{\text{MNS}} 
 (m_{D}^{\nu})^{2} U_{\text{MNS}}^{\dagger}\right)_{ij} .\label{eq:LL:typeII}
\end{equation}

\vspace*{3mm}

\end{document}